%% file: paper.tex
\documentclass[a4paper,USenglish,orivec,runningheads]{llncs}

\input{header}

\begin{document}
\counterwithout{lstlisting}{chapter}

\title{\toolnameUpperCase: A Toolsuite for the\\ Verification of Threshold Automata}

\authorrunning{P. Eichler, T. Baumeister, M. Sakr, M. K. Dowlati, M. V{\"o}lp and S. Jacobs}

\renewcommand{\orcidID}[1]{\,{\href{https://orcid.org/#1}{\includegraphics[scale=0.03]{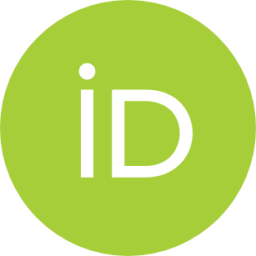}}}}

\author{%
    Paul Eichler\inst{1}\href{mailto:paul.eichler@cispa.de}{\color{black}\Letter}\orcidID{0009-0008-6117-318X}\and
    Tom	Baumeister\inst{1}\orcidID{0009-0009-8539-6246} \and
    Mouhammad Sakr\inst{2}\orcidID{0000-0002-5160-0327} \and
    Mahboubeh Kalateh Dowlati\inst{3} \and
    Marcus V\"olp\inst{3}\orcidID{0000-0002-8020-4446} \and
    Swen Jacobs\inst{1}\orcidID{0000-0002-9051-4050}
}
\institute{
    CISPA Helmholtz Center for Information Security, Germany\\
    \and
    American University of Beirut, Lebanon
    \and
    SnT, Luxembourg University, Luxembourg}

\maketitle

\begin{abstract}
    We present \toolname, a toolsuite for the development and automatic verification of fault-tolerant and threshold-based distributed algorithms.
    Our toolsuite implements three approaches for model checking threshold automata in different decidable fragments known from the literature and two semi-decision procedures going beyond these decidable fragments.
    Moreover, \toolname is a modular, extensible, and well-documented framework for developing algorithms and tools for threshold automata.
    We present important features, give an overview of the implemented algorithms, and evaluate their performance experimentally.
\end{abstract}

\input{intro}

\input{preliminaries}

\input{implementation}

\input{features}

\input{experiments}

\input{small-conclusion.tex}

\begin{credits}
    \subsubsection{\ackname} T. Baumeister and P. Eichler carried out this work as members of the Saarbr\"ucken Graduate School of Computer Science.
    This research was funded in whole or in part by the German Research Foundation (DFG) grant 513487900, by the German Research Foundation (DFG) grant 497132954 and the Luxembourg National Research Fund (FNR) grant C22/IS/17432184. For the purpose of open access, and in fulfilment of the obligations arising from the grant agreement, the author has applied a Creative Commons Attribution 4.0 International (CC BY 4.0) license to any Author Accepted Manuscript version arising from this submission.

    \subsubsection{\discintname}
    The authors have no competing interests to declare that are relevant to the content of this article.

    \subsubsection{Data Availability}
    The benchmarks, evaluation scripts, source code of \toolname and container images used to obtain the results presented in \cref{sec:benchmarks} and \cref{sec:extended-evaluation} are available on Zenodo under Apache 2.0 license at \url{https://doi.org/10.5281/zenodo.19659446}.

    Additionally, the source code and documentation of \toolname is publicly available on GitHub at \url{https://github.com/cispa/TACO} and the documentation is also hosted on the tool website \url{https://taco-mc.dev}.

\end{credits}

\bibliographystyle{splncs04}
\bibliography{reference}
\newpage

\appendix
\input{example}
\input{input-format}
\newpage
\input{extended-evaluation}

\end{document}

%% file: header.tex
\usepackage[T1]{fontenc}

\usepackage{xcolor, colortbl}
\usepackage{amsfonts}
\usepackage{amsmath,thm-restate}
\usepackage{amssymb}

\usepackage{cite}
\usepackage{graphicx}
\usepackage{subfig}
\usepackage[inline,shortlabels]{enumitem}
\usepackage{wrapfig}
\usepackage{environ}
\usepackage{multirow}
\usepackage{xspace}
\usepackage{color}
  \definecolor{lightblue}{rgb}{.8,.95,1}

\usepackage{thmtools, thm-restate}
\usepackage{makecell}
\usepackage{tikz}
\usepackage{pgf}
\usetikzlibrary{positioning,calc,automata,arrows,fit,shapes}

\usepackage{mdframed}

\usepackage{todonotes}
\usepackage{soul}

\usepackage{verbatim}

\usepackage[misc]{ifsym}

\usepackage{lscape}

\usepackage{hyperref}

\usepackage[capitalise,english,nameinlink]{cleveref} 
\crefname{line}{\text{line}}{\text{lines}} %

\newcommand{\mailto}[1]{\href{mailto:#1}{\nolinkurl{#1}}}

\definecolor{mygreen}{RGB}{0,150,0}
\usepackage{pifont} %
\newcommand{\xmark}{\leavevmode{\color{red}\ding{55}}}%
\newcommand{\cmark}{\leavevmode{\color{green!75!black}\ding{51}}}%

\newcommand{\toolname}{\textsc{Taco}\xspace}
\newcommand{\toolnameUpperCase}{TACO}
\newcommand{\bymc}{ByMC~\cite{JohnKSVW13,KonnovVW14,KonnovVW17,KonnovVW15,fmsdKonnovLVW17,KonnovLVW17,0001W18,0001LSW23,bymcgithub}}
\newcommand{\bymcShort}{ByMC~\cite{0001W18,bymcgithub}}

\renewcommand{\vec}[1]{\mathbf{#1}}

\newcommand{\mR}{\mathcal{R}}
\newcommand{\mI}{\mathcal{I}}

\newcommand{\resets}{\tau}
\newcommand{\Shared}{\Gamma}

\newcommand{\vp}{\vec{p}}
\newcommand{\from}{\textit{from}}
\renewcommand{\to}{\textit{to}}
\newcommand{\vup}{\vec{uv}}
\newcommand{\Rules}{\mR}

\newcommand{\vg}{\vec{g}}
\newcommand{\vk}{\vec{k}}

\newcommand{\conf}{\sigma}

\definecolor{darkgreen}{rgb}{0,0.5,0}
\definecolor{darkblue}{rgb}{0,0,.5}
\definecolor{mygray}{gray}{.3}

\newcommand{\state}{q}

\newcommand{\stateset}{\expandafter\MakeUppercase\expandafter{\state}}
\newcommand{\Stateset}{\expandafter\MakeUppercase\expandafter{\State}}

\newcommand{\Nat}{\ensuremath{\mathbb{N}}}
\newcommand{\Int}{\ensuremath{\mathbb{Z}}}

\newcommand{\smartpar}[1]{\medskip \noindent {\bf #1.}}

\newcommand{\li}{\begin{itemize}}
\newcommand{\il}{\end{itemize}}

\tikzstyle{block2}=[draw, text width=9.1em, text centered, minimum height=2.5em, rounded corners, line width = .4mm, font=\bfseries, minimum height = 1.5cm]

\newcommand{\globally}{\square}

\hypersetup{
    colorlinks = true,
    breaklinks = true,
    citecolor   = cispa-dark-blue,
    linkcolor   = cispa-dark-blue,
    urlcolor    = cispa-medium-blue,
}

\usepackage{algorithm}
\usepackage[noend]{algpseudocode}
\algnewcommand{\LineComment}[1]{\State \(\triangleright\) #1}
\usepackage{float}
\newfloat{algorithm}{t}{lop}
\floatname{algorithm}{Algorithm}

\definecolor{cispa-blue}{HTML}{009CC4}
\definecolor{cispa-blue-80}{HTML}{33B0D0}
\definecolor{cispa-blue-60}{HTML}{66C4DC}
\definecolor{cispa-blue-40}{HTML}{99D7E7}
\definecolor{cispa-blue-20}{HTML}{D5ECF6}
\definecolor{cispa-dark-blue}{HTML}{002864}
\definecolor{cispa-medium-blue}{HTML}{018DCA}
\definecolor{cispa-bright-blue}{HTML}{00C8FF}
\definecolor{cispa-anthracite}{HTML}{1C1C1C}
\definecolor{cispa-dark-gray}{HTML}{2B2B2B}
\definecolor{cispa-light-gray}{HTML}{C9C9C9}
\definecolor{cispa-grid-gray-light}{HTML}{E8E8E8}
\definecolor{cispa-offwhite}{HTML}{F5F5F5}
\definecolor{cispa-green}{HTML}{008040}
\definecolor{cispa-red}{HTML}{EB3300}
\definecolor{cispa-orange}{HTML}{FF7F32}
\definecolor{cispa-yellow}{HTML}{FFC600}
\definecolor{cispa-gray}{HTML}{969595}
\definecolor{cispa-blue-10}{HTML}{EBF5F9}
\definecolor{cispa-blue-gray-dark}{HTML}{004C61}
\definecolor{cispa-blue-gray-light}{HTML}{6D8995}
\definecolor{cispa-blue-green}{HTML}{00C49A}
\definecolor{cispa-blue-green-20}{HTML}{DBF4EA}
\definecolor{cispa-blue-green-bright}{HTML}{00EEBB}

\definecolor{bluekeywords}{HTML}{0052CC}
\definecolor{lightbluekeywords}{HTML}{009CC4}
\definecolor{greentypes}{HTML}{00CC44}
\definecolor{graynumbers}{rgb}{0.5, 0.5, 0.5}
\definecolor{greencomments}{rgb}{0.6, 0.4, 0.9}

\definecolor{orange}{RGB}{209, 106, 10}
\definecolor{violet}{RGB}{98, 4, 145}
\definecolor{red}{RGB}{148, 4, 4}

\definecolor{Gray}{gray}{0.925}

\usepackage[]{listings}

\crefname{lstlisting}{Listing}{Listings}

\lstdefinelanguage{ta}{
    morekeywords = [0]{skel, threshAuto, TA, ta},
    keywordstyle=[0]\bfseries\color{lightbluekeywords},
    morekeywords = [1]{parameters, shared, local, assumptions, assume, locations, inits, rules, specifications},
    keywordstyle=[1]\color{bluekeywords},
    morekeywords = [2]{when, do},
    keywordstyle=[2]\color{greentypes},
    comment=[l]{//},
    morecomment=[s]{/*}{*/},
}

\lstdefinelanguage{tla}{
    morekeywords =[0]{MODULE},
    keywordstyle=[0]\bfseries\color{lightbluekeywords},
    morekeywords =[1]{EXTENDS, CONSTANT, ASSUMPTION, VARIABLES, ASSUME },
    keywordstyle=[1]\color{bluekeywords},
    morekeywords =[2]{ EXCEPT, UNCHANGED},
    keywordstyle=[2]\color{greentypes},
    morekeywords =[3]{Cardinality},
    keywordstyle=[3]\color{violet},
    morekeywords =[4]{ TypeOk, Init, Next, Spec},
    keywordstyle=[4]\color{lightbluekeywords},
    comment=[l]{//},
    morecomment=[s]{---}{---},
    morestring=[b]",
}

\usetikzlibrary{arrows,automata,positioning, patterns, patterns.meta} %

\usetikzlibrary{positioning,arrows.meta,decorations.pathmorphing,calc,shapes.multipart,automata}

\tikzstyle{pta}=[auto, ->, >=stealth']
\tikzstyle{every node}=[initial text=]
\tikzstyle{location}=[rectangle, rounded corners, minimum size=12pt, draw=black, fill=blue!10, inner sep=3.5pt]
\tikzstyle{methodBox}=[rectangle, minimum size=12pt, draw=black, fill=green!10, inner sep=3.5pt]
\tikzstyle{invariant}=[draw=black, dotted, fill=yellow, inner sep=1pt, node distance=0] %
\tikzstyle{locguard}=[fill=cyan!30, inner sep=1pt, node distance=0] %
\tikzstyle{final}=[double, fill=blue!50]

\NewEnviron{tikzLTS}{%
  \begin{tikzpicture}[node distance=2cm,inner sep=1pt,minimum size=0.5mm,bend angle=20]
	 	\tikzstyle{proc} = [rectangle,draw=black,fill=green!20,thick,inner sep=10pt]
	  	\tikzstyle{state} = [circle,draw=black,thick,inner sep=3pt]
	  	\tikzstyle{noproc} = [circle]
	  	\tikzstyle{lbl} = [rectangle,node distance=2cm]
  		\tikzstyle{pre} = [ <-,shorten <=2pt,shorten >=2pt, >=stealth', semithick]
	  	\tikzstyle{post} = [ ->,shorten <=2pt,shorten >=2pt, >=stealth', semithick]
    \BODY
  \end{tikzpicture}
}

\NewEnviron{tikzPath}{%
  \begin{tikzpicture}	
  	\tikzstyle{pstate} = [circle, fill=black,thick, inner sep=2pt, minimum size=2mm]
    \tikzstyle{hiddenstate} = [circle]
  	\tikzstyle{trans-notsched} = [ ->,shorten <=2pt,shorten >=2pt, >=stealth', semithick]
  	\tikzstyle{trans-sched} = [ ->,shorten <=2pt,shorten >=2pt, >=stealth', very thick] 
    \BODY
    \end{tikzpicture}
}

\tikzstyle{block}=[
    font=\bfseries,
    line width = .4mm, 
    minimum height = 1.5cm,
]

\tikzstyle{lts}=[auto, ->, >=stealth, font=\small]
\tikzstyle{location}=[
    rectangle,
    rounded corners,
    minimum size=12pt,
    draw=cispa-anthracite,
    fill=cispa-blue-20,
    text=cispa-anthracite,
    inner sep=3.5pt
]

%% file: intro.tex
\section{Introduction}

Ensuring the correctness of the algorithms at the heart of modern distributed systems is a challenging but important task. Many different approaches, models, and tools have been developed to tackle this challenge~\cite{HawblitzelHKLPR17,JaberWJKS21,McMillanP20,RahliGBC15,WilcoxWPTWEA15,Kukovec0W18,PirleaGKZS25}.

Real-world systems require their underlying distributed algorithms to be fault-tolerant and to scale to any number of participants.
While verifying correctness with a parameterized number of processes is already challenging, fault-tolerance additionally requires to reason modulo a \emph{resilience condition} that bounds the type and number of faults that can be tolerated.
Without such a condition, it is impossible for many essential algorithms to achieve their desired properties~\cite{LamportSP82}.

In \toolname, we have chosen threshold automata (TA) as the underlying system model as it supports a parametric number of processes as well as resillience conditions~\cite{KonnovVW14,KonnovVW17,Kukovec0W18,0001LSW23}.
This model allows expressing many essential consensus algorithms~\cite{BrasileiroGMR01,MostefaouiMPR03,BrachaT85,SPIN13,BertrandKLW19}, as well as more advanced distributed algorithms, e.g., from modern blockchain applications~\cite{CrainNG21,0001LSW23}.
Intuitively, a TA models the local state of a participant of a protocol in the automaton locations and the global state by shared integer variables.
These variables essentially count how often a given event (e.g., broadcast of a certain message) has happened so far.
Thresholds can be assigned as guards to transitions of the TA, such that an action can only be taken after the threshold is crossed.

Moreover, different fragments of TA have been shown to support decidable \emph{parameterized verification}, i.e., their correctness can be decided independent of the number of participants~\cite{JohnKSVW13,KonnovVW14,KonnovVW15,fmsdKonnovLVW17,KonnovLVW17,KonnovLVW16,Kukovec0W18,0001W18,BertrandKLW19,Balasubramanian20,Stoilkovska0WZ20,StoilkovskaKWZ21}.
As the restrictions of these decidable fragments may be too strict for certain applications, such as round-based algorithms with resets of shared variables, more generalized verification approaches based on semi-decision procedures have also been considered~\cite{BaumeisterEJSV24}.

While TA can be used to express and verify many interesting distributed algorithms, and the \emph{theory} of TA is well-documented and steadily growing, the situation is not as satisfactory on the tool side:
The only publicly available tool for TA is \bymc, which is no longer maintained.
Implementations of other decision procedures for TA and variants of TA are prototypes not well-suited for use by non-expert users and further development by third parties~\cite{Balasubramanian20,balagit,StoilkovskaKWZ19,StoilkovskaKWZ22,StoilkovskaKWZ21,BaumeisterEJSV24}.
The only related tool that is publicly available and currently supported is PyLTA, a model checker for layered threshold automata (LTA)~\cite{ThomasS23}.
However, LTA are not a generalization of standard TA, and PyLTA does not support the existing TA benchmark library.

\smartpar{Contributions}
In this paper, we present \toolname\footnote{TACO's source code is available on GitHub \url{https://github.com/cispa/TACO} and the documentation is on the webpage \url{https://taco-mc.dev}. A reproduction package containing the source code, benchmark files and documentation at the time of submission is available on Zenodo~\cite{tacoartifact}.}, a toolsuite for the development and verification of threshold-based distributed algorithms.
\toolname provides:
\begin{enumerate}[noitemsep,topsep=5pt]
\item an efficient implementation of three model checking algorithms that scale to previously unsolved TA and extensions of the standard TA model, and
\item a documented, maintained, and extensible code base, well-suited to serve as the basis of future model checkers and tools for TA, as well as
\item an intuitive user interface that supports users in modeling and debugging distributed algorithms.
\end{enumerate}

In this paper, we will give a high-level overview of the three algorithms, describe important features of \toolname, and evaluate the performance of the algorithms on different classes of benchmarks, comparing their strengths and weaknesses.

%% file: preliminaries.tex
\section{System Model and Specifications}\label{sec:preliminaries}
We first define the underlying formal model of TA and introduce the specification language, before providing more details on the architecture of \toolname.

\begin{definition}[Threshold Automaton~\cite{KonnovVW14,KonnovVW17}]\label{def:TA}
A \emph{threshold automaton} (TA) is a tuple $A=(L, \mI, \Shared, \Pi,\Rules, RC)$ where $L$ is a finite set of \emph{locations}, $\mI \subseteq L$ the set of \emph{initial locations}, $\Shared$ is a finite set of \emph{shared variables} over $\Nat_0$, $\Pi$ is a finite set of \emph{parameter variables} over $\Nat_0$, $RC$ is the \emph{resilience condition}, a linear integer arithmetic formula over parameter variables, and $\Rules$ is a set of rules.
A \emph{rule} is a tuple $r=(\from, \to, \varphi,\vup)$ where, $\from,\to \in L$, $\vup \in \Nat_0^{|\Shared|}$ the \emph{update vector} for shared variables and $\varphi$ is a conjunction of lower guards and upper guards.
A \emph{lower guard} is a constraint $a_0 {+} \sum_{i=1}^{|\Pi|} a_i {\cdot} p_i \leq x$, with $x \in \Shared$, $a_0,\ldots,a_{|\Pi|} \in \mathbb{Q}$, $p_1,\ldots,p_{|\Pi|} \in \Pi$.
An \emph{upper guard} is a constraint $ a_0 {+} \sum_{i=1}^{|\Pi|} a_i {\cdot} p_i > x$.	
The left-hand side of a lower or upper guard is a \emph{threshold}. 
\end{definition}

Note that the update vector $\vup$ is in $\Nat_0^{|\Shared|}$ and consequently shared variables can only increase. In the following, we will also call such an automaton a \emph{monotonic threshold automaton} (MTA).

MTA do not allow variable decrements or resets, which restricts their ability to model algorithms, including round-based distributed algorithms, whose execution relies on resetting or reusing rounds.
To address this limitation, \toolname additionally supports \emph{extended threshold automata} (ETA)~\cite{BaumeisterEJSV24}, where $\vup \in \Int_0^{|\Shared|}$, i.e., variables can also be decremented, and a rule contains an additional set $\resets \subseteq \Shared$ which contains the variables that are reset to 0.

To simplify the presentation, where the distinction does not affect our results, we refer to both models simply as threshold automata (TA). Otherwise, we distinguish between ETA and MTA as required.

Common parameter variables in TA are $\Pi = \{n,t,f\}$, where $n$ is the number of processes, $t$ is a bound on the number of faulty processes, and $f$ is the actual number of faulty processes.
This allows $RC$ to express assumptions about the fraction of faulty processes in the system, e.g., $RC = n > 3t\, \land\, t \geq f$.

\input{example_cav}

\smartpar{Configurations} 
A \emph{configuration} of a TA is a triple $\sigma = (\vk, \vg, \vp)$, where $\vk \in \Nat_0^{|L|}$ assigns a number of processes to each location, $\vg \in \Nat_0^{|\Shared|}$ is a valuation of the shared variables, and $\vp$ is a parameter valuation such that $\vp \models RC$, i.e., $RC$ holds after substituting each parameter with its value given by $\vp$.
We say that a configuration is \emph{initial} if $\vg = \vec{0}$ and $\forall i$ $\vk[i] > 0 \Rightarrow l_i \in \mI$.

\smartpar{Paths}
A rule $r = (\from, \to, \varphi,\vup,\resets)$ is \emph{enabled} in a configuration $\sigma = (\vk,\vg,\vp)$ if $\vec{k}[\from] > 0$ and $(\vg,\vp) \models \varphi$, producing $\sigma' = r(\sigma)$, where $r(\sigma)$ denotes the resulting configuration after some process executes $r$, updating $\vk$ and $\vg$.
A \emph{path} $\pi$ is then a sequence of configurations $\pi = \sigma_0, \sigma_1, \ldots$ such that for each $i$ there exists an enabled rule $r_i$ with $\sigma_{i+1} = r_i(\sigma_i)$.
A configuration $\conf_r$ is said to be reachable if there exists a path starting from an initial configuration and ending in $\conf_r$.
We also refer to a path reaching an error configuration as an \emph{error path}.

\smartpar{Specifications}
\toolname supports safety properties in the temporal logic $ELTL_{FT}$ \cite{KonnovLVW17}, which can express reachability of configurations where certain locations are empty or non-empty.
Note that, in contrast to \bymcShort, \toolname currently does not support liveness properties.

However, \toolname allows lower integer bounds on the number of processes in a location.
Formally, the supported specification syntax is defined as follows:
\begin{align*}
\psi &::= \mathsf{pform} \;\;|\;\; \globally \psi \;\;|\;\; \psi \wedge \psi \\
\mathsf{pform} &::= \mathsf{cform} \;\;|\;\; \mathsf{gform} \vee \mathsf{cform} \\
\mathsf{cform} &::= \conf.\vk[l] * n 
        \;\;|\;\; \conf.\vk[l] = 0 
        \;\;|\;\; \conf.\vk[l] \neq 0 
        \;\;|\;\; \mathsf{cform} \wedge \mathsf{cform}
        \;\;|\;\; \mathsf{cform} \vee \mathsf{cform}\\
\mathsf{gform} &::= \varphi
       \;\;|\;\; \neg \mathsf{gform}
       \;\;|\;\; \mathsf{gform} \wedge \mathsf{gform}
\end{align*}

\noindent
where $\varphi$ is a rule guard as defined in \cref{def:TA}, $l \in L, n \in \mathbb{N}$, $* \in \{>, \geq\}$, and $\globally$ denotes the temporal operator 'globally'.
For the example above, the specification $\globally \left( \conf.\vk[d_0] = 0 \vee \conf.\vk[d_1] = 0 \right)$ states that processes cannot decide on different values. 

\smartpar{Parameterized Verification}
More formally, the problem that \toolname aims to solve is checking these specifications against the \emph{parameterized system} specified by the TA, i.e., the infinite family of concrete transition systems resulting from instantiating parameter variables with valuations that satisfy the resilience condition $RC$.
This problem is known to be decidable for MTA~\cite{KonnovVW14,Balasubramanian20,BaumeisterEJSV24} and undecidable for ETA (as a consequence of results in~\cite{Kukovec0W18}).

%% file: example_cav.tex
\input{resources/floodmin-alg-ta-cav.tex}
\begin{example}
\cref{fig:alg-floodmin} shows pseudocode for a naive voting protocol.
Figure~\ref{fig:ta-floodmin} sketches the corresponding TA (taken from \cite{BaumeisterEJSV24}) with $\mI=\{v_0,v_1\}$, $L = \{v_0,v_1,$ $\textit{wait}, d_0,d_1\}$, $\Shared=\{x_0,x_1\}$, $\Pi=\{n,t\}$. 
		A process in $v_0$ has a vote of $0$ and a process in $v_1$ has a vote of $1$. 
		If at least $n-t$ processes vote with $0$~(respectively, $1$), the decision will be $0$~($1$), modeled by all processes moving to $d_0$ ($d_1$).
		
For a more advanced example, including variable resets, see Appendix~\ref{apx:rbp}.
\end{example}
\vspace{-0.2cm}

%% file: resources/floodmin-alg-ta-cav.tex
\vspace{-1cm} 
 \begin{figure}[h]
  \centering
  \begin{minipage}[t]{0.49\textwidth}
  \vspace{0pt} 
    \centering
    \begin{algorithm}[H]
        \caption{}
        \begin{algorithmic}[1]
        \State \textbf{int} $v \gets \text{input}(\{0, 1\})$
        \State broadcast $v$;
        \State $R \gets $ receive messages from all;
            \If{$|\{ r \in R : r = 1 \}| > n - t$}
                \State decide $1$;
            \EndIf
            \If{$|\{ r \in R : r = 0 \}| > n - t$}
                \State decide $0$;
            \EndIf
        
        \end{algorithmic}
        \end{algorithm}
        \vspace{-25pt}
    \caption{A simple voting protocol.}
    \label{fig:alg-floodmin}
  \end{minipage}
  \hfill
  \begin{minipage}[t]{0.49\textwidth}
    \vspace{41.75pt} 
    \scalebox{.80}{
           \begin{tikzpicture}[thick, lts, font=\scriptsize]		
                \node [location, initial, initial text=] (v0) at (0,0) {$v_0$};
                \node [location] (w)  at (3, -1) {\textit{wait}};
                \node [location, initial, initial text=] (v1) at (0,-2)  {$v_1$};
                \node [location] (d0) at (6,0)  {$d_0$};
                \node [location] (d1)  at (6,-2)  {$d_1$};

                \path[->]
                    (v0) edge [] node [sloped, above] {$r_0: x_{0}++$} (w)
                    (v1) edge [] node [sloped, below] {$r_1: x_{1}++$} (w)
                    (w) edge [] node [sloped, above] {$r_2: x_{0} \geq n-t$} (d0)
                    (w) edge [] node [sloped, below] {$r_3: x_{1} \geq n-t$} (d1)
                ;
		\end{tikzpicture}
    }%
    \vspace{5.5pt}
    \caption{TA of the algorithm in \cref{fig:alg-floodmin}}
    \label{fig:ta-floodmin}
  \end{minipage}
\end{figure}
\vspace{-0.5cm} 

%% file: implementation.tex
\section{Architecture and Implementation}
\label{sec:implementation}

A high-level overview of the architecture of \toolname is provided in \cref{fig:architecture}.
At the heart of \toolname are three different model checking algorithms, based on representations for different flavors of TA, on different abstraction levels.
Additionally, it contains a parser for specification files, a preprocessor, unified interfaces to binary decision diagram (BDD)~\cite{bdds} libraries and satisfiability modulo theories (SMT) solvers, as well as utility code (not shown in the figure), e.g., for the command line interface (CLI).
All components are designed to be usable individually to enable fast and easy development of, for example, new model checking algorithms, different user interfaces, new specification formats or testing of new backends like a new BDD library.

We describe the most interesting components in more detail below.
For algorithmic details of the model checking algorithms, we refer the reader to the respective original publications~\cite{Balasubramanian20,BaumeisterEJSV24}.

\input{./resources/architecture-diagram.tex}

\smartpar{Tool-Supported Modelling Languages}
\toolname supports models in two different languages. 
The first is an extension of the language introduced by ByMC for modelling threshold automata~\cite{0001W18,bymcgithub,benchmarksgithub}, which are usually denoted by the file ending \texttt{.ta} (\texttt{.eta} for extended threshold automata).
The second is a fragment of TLA$^+$~\cite{Lamport2002}, a more well-known language for model checking, which enables expressing the same models in a logic-based framework. 
The grammars of both languages are available in the tool repository
 and an informal description of the formats on the tool page.
A \texttt{.ta} model corresponding to the threshold automaton depicted in \cref{fig:ta-floodmin} is provided in \cref{lst:alg-simple} and \cref{apx:tla} provides the same model written in the TLA$^+$ fragment.

\vspace{5mm}
\begin{lstlisting}[caption={TA of \cref{fig:ta-floodmin} encoded as a \texttt{.ta} file},captionpos=b, label=lst:alg-simple]
ta ALG1 {
    shared x0, x1;
    parameters n, t;
    assumptions (2) { n > 3 * t; }
    locations (5) { V0: [0]; V1: [1]; D0: [2]; D1: [3]; WAIT: [4]; }
    inits(4) {  WAIT == 0; D0 == 0; D1 == 0; V1 + V0 == n; x0 == 0; x1 == 0; }
    rules (4) {
        0: V0 -> WAIT when (true) do { x0' := x0 + 1; };
        1: V1 -> WAIT when (true) do { x1' := x1 + 1;  };
        2: WAIT -> D0 when (x0 >= n - t) do {};
        3: WAIT -> D1 when (x1 >= n - t) do {};
    }
    specifications (1) { 
        cor: [](!(D0 > 0 && D1 > 0))
     }
}
\end{lstlisting}

\smartpar{Preprocessor}
The \toolname preprocessor parses the TA and performs a sanity check on the input, followed by a static analysis of the TA and the specification in order to simplify the problem.
This includes removing self-loops without effect on the variables
and transition guards that are always satisfied under the given resilience condition of the TA.
Optionally, the preprocessor can also remove transitions with guards that will never be satisfied in the given TA.
Moreover, if \toolname can statically determine that, under the assumptions on initial states given in the specification, some locations are never reachable, then these locations are completely removed before checking the property.

\smartpar{SMT Model Checking Algorithm}
Balasubramanian et al.~\cite{Balasubramanian20} showed that the reachability problem for MTA can be reduced to deciding the satisfiability of an SMT formula.
More precisely, they construct an existential Presburger arithmetic formula $\phi_{REACH}(\sigma,\sigma')$ such that an assignment for $\sigma, \sigma'$ satisfies $\phi_{REACH}(\sigma,\sigma')$ if and only if $\sigma'$ is reachable from $\sigma$ for the given MTA.
This algorithm directly uses a basic TA representation shown as LIATA in \cref{fig:architecture}, corresponding closely to the definition of ETA.

In contrast to the original implementation described in~\cite{Balasubramanian20} and published later at~\cite{balagit}, we do not use non-deterministic guessing to keep SMT queries small.
Instead, we encode all steady paths and leave it to the SMT solver to resolve the non-determinism.
Note that this algorithm can only verify correctness of MTA since it relies on the monotonicity of guards, which requires that whenever a lower (upper) guard is disabled (enabled), it stays disabled (enabled) forever.

\smartpar{ZCS Model Checking Algorithm}
This algorithm, originally described in~\cite{BaumeisterEJSV24}, is based on an abstraction of the parameterized system semantics to a set of \emph{$01$-counter systems}~(ZCS)~\cite{JohnKSVW13,PnueliXZ02,EichlerJW25}.
It uses the IntervalTA representation, which is derived from LIATA, where the main difference is that shared variables are mapped to a finite domain through \emph{parametric interval abstraction}~\cite{JohnKSVW13,KestenP00}.
This interval abstraction depends on a total ordering between thresholds.
As this may depend on parameter valuations, \toolname uses the SMT solver to compute all possible orders under the resilience condition, and then constructs the corresponding IntervalTA and the ZCS representing its abstract semantics for each of them.%
The finite state space of each ZCS is represented and manipulated using BDDs.
Based on a set of ZCS error configurations, derived from the specification, the ZCS Model Checker performs a BDD-based backward exploration from the error configurations until reaching a fixed point.
Finally, it checks breadth-first if the resulting error graph contains a non-spurious path, i.e., a path of the ZCS that can be instantiated to a concrete error path.
Note that for ETA, the procedure is only a semi-decision procedure.%

\smartpar{ACS Model Checking Algorithm}
\noindent The ACS Model Checker follows a similar workflow as the ZCS Model Checker, but operates on an \emph{abstract counter system} (ACS) that keeps track of the concrete number of processes in each location.
It was shown in~\cite{BaumeisterEJSV24} that an ACS forms a WSTS, which enables parameterized safety verification by a fixed point computation using finite representations of infinite sets of configurations~\cite{Finkel87,AbdullaCJT96}.

The ACS Model Checker operates on the ACSTA representation, which is derived from the IntervalTA by amending it with a representation of configurations that enables efficient implementation of the WSTS-based fixed point computation. 
Based on this representation, the algorithm maintains a representation of all visited configurations in a tree data structure, facilitating fast lookups of comparable configurations (with respect to the well-quasi order (wqo)).
As a heuristic that has proven useful, the SMT checks for spuriosity of abstract error paths are done incrementally over the length of the path.
This allows us to exclude whole sets of abstract paths with relatively simple SMT queries, and the SMT solver can re-use information from the shorter paths when checking their extensions.
However, the performance of this heuristic depends on how efficiently the SMT solver supports incremental queries.
Note that for ETA, the procedure is only a semi-decision procedure and does not support specifications that require target locations to be empty, for more details refer to \cite{BaumeisterEJSV24}.

%% file: resources/architecture-diagram.tex
\tikzstyle{component}=[
    fill=white, 
    rounded corners, 
    line width=0.5mm, 
    text centered, 
    minimum height=0.6cm, 
    minimum width=1cm,
    text width=18mm,
]   

\definecolor{spec}{RGB}{255, 179, 102}
\definecolor{preprocessor}{rgb}{0.6, 0.4, 0.9}
\definecolor{ta1}{HTML}{0052CC}
\definecolor{mc1}{HTML}{00CC44}
\definecolor{smt-encoder}{rgb}{0.2, 0.4, 0.6}
\definecolor{res}{rgb}{0.6, 0.2, 0.5}

\definecolor{ext}{HTML}{009CC4}
\tikzstyle{ext-component}=[
    draw=ext,
    fill=ext!10, 
    line width=0.2mm, 
    text centered, 
    minimum height=5.4mm, 
    minimum width=7mm,
    text width=6mm,
    font=\tiny
]   

\begin{figure}[h]
    \centering
    \scalebox{0.9}{
        \hspace{-2.5mm}

        \begin{tikzpicture}[node distance=1cm and 1.5cm, auto, thick, font=\scriptsize, >=stealth]
            \node[component, draw=spec, fill=spec!10] at (0,0) (input) {Input\\\texttt{.ta\,/.eta\,/.tla}};

            \node[component, draw=preprocessor, fill=preprocessor!10] at (0,-1) (preproc) {Preprocessor};
            \path[->]
                (input) edge [] node {} (preproc)
            ;

            \node[component, draw=ta1, fill=ta1!15] at (0,-2.5) (lta) {LIATA};
            \node[component, draw=ta1, fill=ta1!15] at (0,-4) (ita) {IntervalTA};
            \node[component, draw=ta1, fill=ta1!15] at (0,-5.5) (ata) {ACSTA};
            \path[->]
                (preproc) edge [] node [yshift=1.75mm] {parse} (lta)
                (lta) edge [] node {derive} (ita)
                (ita) edge [] node {derive} (ata)
            ;

            \draw[-, dashed, draw=gray, fill=gray, fill opacity = 0.1]
                (-1.3,-1.8) -- (1.3, -1.8) 
                -- (1.3, -6.2) -- node [text width=26mm, fill opacity=1] {\vspace{-3.5mm}\begin{center}Threshold Automata\\ Representations\end{center}} (-1.3, -6.2)
                -- (-1.3,-1.8)
            ;

            \node[component, draw=mc1, fill=mc1!15] at (4, -2.5) (smtmc) {SMT Model Checker};
            \node[component, draw=mc1, fill=mc1!15] at (4, -4) (zcsmc) {ZCS Model Checker};
            \node[component, draw=mc1, fill=mc1!15] at (4, -5.5) (acsmc) {ACS Model Checker};
            \path[->]
                (lta) edge [] node {input} (smtmc)
                (ita) edge [] node {input} (zcsmc)
                (ata) edge [] node {input} (acsmc)
            ;   

            \draw[-, dashed, draw=gray, fill=gray, fill opacity = 0.1]
                (2.7, -1.8) -- (5.3, -1.8) 
                -- (5.3, -6.2) -- node [yshift=-0.75mm, fill opacity=1] {Model Checkers} (2.7,-6.2)
                -- (2.7, -1.8)
            ;

            \node[component, draw=smt-encoder, fill=smt-encoder!10] at (7.5, -4) (encoder) {SMT Encoder};
            \path[-]
                (smtmc) edge [above] node {all paths} (7.5, -2.5)
                (acsmc) edge [below] node {error paths} (7.5, -5.5)
            ;
            \path[->]
                (7.5, -2.5) edge [] node {} (encoder)
                (zcsmc) edge [above] node [yshift=-3.75mm, xshift=2mm,  text width=10mm] {error\\paths} (encoder)
                (7.5, -5.5) edge  [] node {} (encoder)
            ;

            \node[component, draw=res, fill=res!10] at (10.5, -4) (result) {Result};
            \path[->]
                (encoder) edge [] node {} (result)
            ;

            \node[ext-component] at (6, -3.2) (bdd) {BDD};
            
            \node[ext-component] at (9, -3.2) (solver) {SMT Solver};
            \draw[->, dotted]
                (bdd)  -- (4,-3.2) -- (zcsmc);
            \draw[->, dotted]
                (solver) -- (7.9, -3.2) -- (encoder.39);
        \end{tikzpicture}
    }
    \caption{Architecture of the TACO toolsuite}
    \label{fig:architecture}
\end{figure}
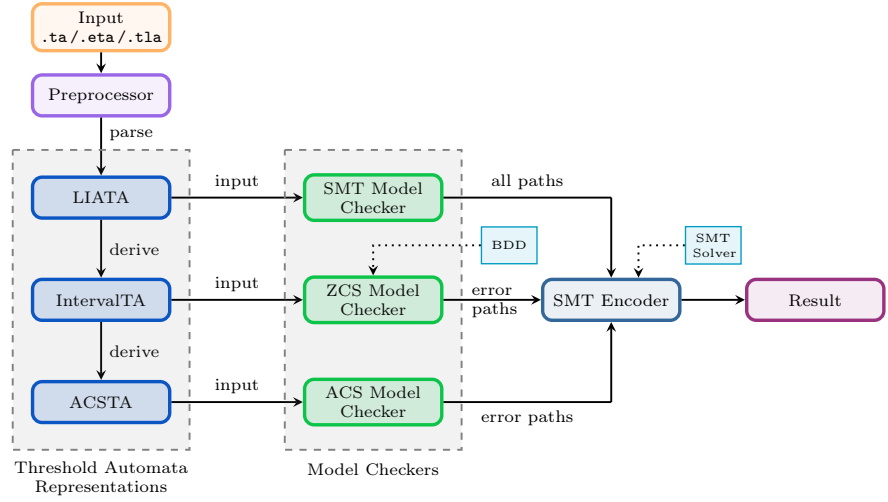

%% file: features.tex
\section{Tool Features}\label{sec:features}

\smartpar{Basic User Experience}
\toolname supports the fully automatic verification of threshold automata: Supply it with a TA and a specification, and it will automatically choose a model checking algorithm\footnote{The default model checker for standard TA is the SMT model checker, and for ETA the ZCS model checker.} and answer whether the specification is satisfied.
This allows users who are not experts to use the tool, as long as they can produce inputs in one of the supported formats.

The basic input format for TA is adopted from \bymcShort, enabling intuitive modeling and re-use of the existing benchmark library.
In addition, \toolname supports a fragment of TLA$^+$, which is more welcoming for users familiar with TLA$^+$ and also allows checking fixed-size instances of the protocol.

Both formats allow the inclusion of specifications in $ELTL_{FT}$ (described in \cref{sec:preliminaries}).

\toolname has been developed with user experience in mind.
In case of user errors such as faulty inputs, it provides easy-to-read and extensively documented API and CLI interfaces.

\smartpar{Advanced Features}
Although \toolname can be used as a push-button tool, it offers lots of additional features for advanced users.
Besides a simple yes/no answer on whether the current protocol satisfies the specification, \toolname also provides algorithm designers with richer, informative outputs.
This includes the visualization of TA for more intuitive modeling and verification, concrete error traces if an error is found during verification, or intermediate results such as abstract error paths.
These outputs can help the designer to identify unintended behavior of the protocol and amend it accordingly.

Additionally \toolname exposes a wealth of configuration parameters for experts.
The user can, for example, choose which of the three model checking algorithms to use or configure the backend components. This includes configuring SMT solver tactics or setting the reordering method of the BDD managers.

\smartpar{Extensions of the System Model}
In contrast to previous tools for the verification of threshold automata, \toolname goes beyond the decidable fragment.
It is the first tool to support ETA.
While \toolname is not guaranteed to terminate for such inputs, our experiments (in \cref{sec:benchmarks}) show that it can solve a number of benchmarks from this class.%

\smartpar{Code Quality and Modularity}%
The goal of a model checker is to verify the correctness. 
However, to meaningfully increase the confidence in a protocol, the verifier must be trustworthy itself.
Therefore, we use extensive testing, achieving $\sim97\%$ line coverage at the time of writing, and software engineering best practices such as code reviews and linting to maintain a high-quality code base.
Additionally, \toolname is written entirely in the safe fragment of Rust, guaranteeing memory and type safety while not adding major overhead and allowing for low-level optimizations.

Moreover, to enable others to write tools for TA based on \toolname, we designed it in a highly modular and reusable fashion.
Most major components (e.g., parser, preprocessor, TA representations, and model checking algorithms) are separated into individual, publicly available Rust crates. 
This should allow others to easily re-use them and, for example, add other model checking algorithms.

The same modular approach has been taken for backend components such as SMT solvers or BDD libraries.
\toolname supports any SMT solver that uses the SMTLIB2 standard~\cite{barrett2010smt} and has an interactive mode.
For BDD libraries we implemented a general high-level interface, which requires minimal code to add a new library, and we currently support CUDD~\cite{cudd,Somenzi01} and OxiDD~\cite{oxidd24}.

%% file: experiments.tex
\section{Experimental Evaluation}
\label{sec:benchmarks}

We evaluate all three of our model checking algorithms and compare them against the only available existing tool for verification of TA, \bymcShort. 
Other implementations of approaches from the literature, such as Balasubramanian et al.~\cite{Balasubramanian20,balagit} or Baumeister et al.~\cite{BaumeisterEJSV24}, are research prototypes that do not support our benchmark suite without additional manual effort, for example manual input for the parametric interval abstraction~\cite{BaumeisterEJSV24} or encoding of benchmarks by hand into internal data structures~\cite{Balasubramanian20,balagit}.
Additionally, comparing \toolname to the execution times reported in \cite{BaumeisterEJSV24}, \toolname significantly outperforms the implementation in \cite{BaumeisterEJSV24}, especially on the larger ETA benchmarks.

\smartpar{Setup} All model checkers were run in their default settings\footnote{We also attempted to run an additional mode of ByMC but it reported invalid counterexamples, for more details, see \cref{sec:exclusion-cav15}.} and set to terminate upon detecting a violated property. By default, \toolname uses Z3~\cite{z3} as SMT solver, and the ZCS model checker uses CUDD~\cite{cudd,Somenzi01} as the BDD library.

The experiments were conducted on machines with two AMD EPYC 7773x processors, with 64 cores each and 2TB RAM. 
Time and memory consumption was monitored with the \texttt{time} command and we report the elapsed wall-clock time. A more detailed description of the setup can be found in \cref{sec:evaluation-environment}.

\smartpar{Benchmark Sets}
Overall, we used five sets of benchmarks for our evaluation.
For MTA (supported by all model checkers), we used
\begin{enumerate*}
    \item \emph{ByMC Handcoded ISOLA18}, a set of handcrafted TA that appeared in \cite{0001W18},
    \item \emph{ByMC ISOLA18 Promela}, a set of benchmarks that was obtained by executing the Promela translation of ByMC and also from~\cite{0001W18}, and
    \item \emph{RedBelly small,} containing Red Belly blockchain components \cite{CrainNG21} modeled as TA \cite{holistic,BaumeisterEJSV24}.
\end{enumerate*}
\footnote{
    For more details on the origin and selection of benchmarks, refer to \cref{sec:origin-benchmarks}.
}
As mentioned in \cref{sec:preliminaries}, \toolname only supports the safety fragment of $ELTL_{FT}$. Therefore, the comparisons were performed on properties within that fragment.

Additionally, we compared the model checkers that support ETA (our ZCS and ACS model checkers) on two variants of a multi-phase King Consensus protocol and a set of multi-phase RedBelly benchmarks taken from~\cite{BaumeisterEJSV24}.

The execution times and memory consumption for all model checkers are given in \cref{tab:benchmark-ta}. \footnote{Extended benchmark results ,e.g., \toolname with CVC5, can be found in \cref{sec:extended-eval-results}.}

\newcolumntype{g}{>{\columncolor{Gray}}c}

\begin{table}
    \caption{Benchmark results for \toolname and ByMC. Columns $|L|$ and $|\Rules|$ give the number of locations and rules of the TA, respectively. In column ``safe?'' \cmark\xspace denotes that all checked properties hold, \xmark\xspace  denotes a violated property was found, and $?$ that the result is unknown. Execution time (here wall clock time) is given in seconds, \emph{TO} denotes a timeout after 20min. The fastest execution time per benchmark is highlighted in \textbf{bold font}. RSS columns give the maximal resident set size in GB. ``Err'' denotes that an error was returned.}
    \label{tab:benchmark-ta}
    \begin{center}
        \scalebox{0.78}{
            \begin{tabular}{|l | g c g || r g | r g | r g | r g|}
                \hline
                \multirow{3}{*}{\textbf{Benchmark}} & \cellcolor{white} & & \cellcolor{white} & & \cellcolor{white} & \multicolumn{6}{c|}{\textbf{TACO}} \\
                \cline{7-12}
                 & \multicolumn{3}{c||}{\multirow{-2}{*}{\textbf{TA}}} & \multicolumn{2}{c|}{\multirow{-2}{*}{\textbf{ByMC}}} & \multicolumn{2}{c|}{\textbf{SMT}} & \multicolumn{2}{c|}{\textbf{ACS}} & \multicolumn{2}{c|}{\textbf{ZCS}} \\
                 & $|L|$ & $|\Rules|$ & safe? & time & RSS & time & RSS & time & RSS & time & RSS  \\
                \hline\hline
                \textbf{MTA Benchmarks} & & & & & & & & & & & \\
                \hline
                \textbf{ByMC Handcoded ISOLA18\cite{0001W18}} & & & & & & & & & & & \\
                    aba&5&10&\cmark&0.69&0.04&0.07&0.03&\textbf{0.06}&\textbf{0.03}&0.07&0.03 \\
                    bcrb&5&13&\cmark&0.42&0.04&0.08&0.03&\textbf{0.07}&\textbf{0.03}&\textbf{0.07}&\textbf{0.03} \\
                    bosco&8&20&\cmark&139.33&0.07&\textbf{1.03}&\textbf{0.04}&117.29&0.04&160.34&0.05 \\
                    c1cs&9&30&\cmark&TO&-&\textbf{0.72}&\textbf{0.05}&1.68&0.03&29.25&0.05 \\
                    cc&7&14&\cmark&0.40&0.04&\textbf{0.26}&\textbf{0.04}&TO&-&TO&- \\
                    cf1s&9&26&\cmark&397.75&0.10&0.22&0.04&\textbf{0.17}&\textbf{0.03}&0.28&0.05 \\
                    nbacg&8&16&\cmark&0.36&0.04&\textbf{0.24}&\textbf{0.05}&0.24&0.05&0.41&0.05  \\
                    nbacr&7&16&\cmark&0.31&0.04&\textbf{0.08}&\textbf{0.05}&\textbf{0.08}&\textbf{0.05}&0.11&0.05 \\
                    strb&4&8&\cmark&0.26&0.04&0.06&0.03&0.06&0.03&\textbf{0.04}&\textbf{0.03} \\
                \hline
                \textbf{RedBelly small\cite{BaumeisterEJSV24}} & & & & & & & & & & & \\
                rb-bc&10&19&\cmark&0.96&0.04&0.14&0.03&\textbf{0.09}&\textbf{0.03}&0.11&0.03\\
                rb-simple&19&33&\cmark&TO&-&1.24&0.07&7.18&0.82&\textbf{0.81}&\textbf{0.03}\\
                rb&26&41&\cmark&TO&-&10.04&0.49&58.47&3.44&\textbf{0.85}&\textbf{0.03}\\
                \hline
                \textbf{ByMC ISOLA18 Promela\cite{0001W18}} & & & & & & & & & & & \\
                aba\_case1&37&202&\cmark&15.19&0.06&2.08&0.17&\textbf{1.19}&\textbf{0.03}&1.43&0.04 \\
                aba\_case2&61&425&\cmark&290.42&0.11&19.49&0.74&17.79&0.09&\textbf{15.12}&\textbf{0.15} \\
                bosco\_case1&28&152&\xmark&45.16&0.06&\textbf{7.61}&\textbf{0.09}&332.93&0.06&104.76&0.19 \\
                bosco\_case2&40&242&\xmark&752.65&0.08&\textbf{38.47}&\textbf{0.21}&TO&-&TO&- \\
                bosco\_case3&32&188&\xmark&52.14&0.06&\textbf{13.30}&\textbf{0.12}&982.39&0.10&117.45&0.17 \\
                c1cs\_case1&101&1285&\xmark&Err&-&TO&-&TO&-&\textbf{83.00}&\textbf{0.22} \\
                c1cs\_case2&70&650&\xmark&TO&-&1164.87&14.06&102.70&0.41&\textbf{26.23}&\textbf{0.15} \\
                c1cs\_case3&101&1333&\xmark&TO&-&TO&-&TO&-&TO&- \\
                cc\_case1&164&2064&?&Err&-&TO&-&TO&-&TO&- \\
                cc\_case2&73&470&\xmark&Err&-&\textbf{1178.05}&\textbf{8.82}&TO&-&TO&- \\
                cc\_case3&304&6928&?&Err&-&TO&-&TO&-&TO&- \\
                cc\_case4&161&2105&?&Err&-&TO&-&TO&-&TO&- \\
                cf1s\_case1&41&280&\cmark&18.31&0.07&8.22&0.90&\textbf{0.43}&\textbf{0.03}&0.59&0.03 \\
                cf1s\_case2&41&280&\cmark&102.08&0.10&7.25&0.90&\textbf{0.54}&\textbf{0.03}&0.90&0.03 \\
                cf1s\_case3&68&696&\cmark&TO&-&41.35&1.17&\textbf{7.99}&\textbf{0.28}&8.53&0.09 \\
                frb&7&14&\cmark&0.36&0.03&\textbf{0.06}&\textbf{0.03}&\textbf{0.06}&\textbf{0.03}&\textbf{0.06}&\textbf{0.03} \\
                nbacg&24&64&\xmark&\textbf{0.30}&\textbf{0.04}&0.40&0.06&0.41&0.05&0.41&0.05 \\
                nbacr&77&1031&\xmark&\textbf{1.69}&\textbf{0.17}&TO&-&TO&-&8.65&0.22 \\
                strb&7&21&\cmark&0.34&0.04&\textbf{0.07}&\textbf{0.03}&\textbf{0.07}&\textbf{0.03}&\textbf{0.07}&\textbf{0.03} \\
                \hline\hline 
                \textbf{ETA Benchmarks} & & & & & & & & & & & \\
                \hline
                \textbf{King Consensus}\cite{BaumeisterEJSV24} & & & & & & & & & & & \\
                phase-king-buggy&27&10&\xmark& & & & &\textbf{48.84}&\textbf{3.23}&TO&- \\
                phase-king&27&10&\xmark& & & & &\textbf{64.21}&\textbf{3.73}&410.01&0.38 \\
                \hline
                \textbf{RedBelly with resets}\cite{BaumeisterEJSV24} & & & & & & & & & & & \\
                rb-2x\_reset&49&28&\cmark& & & & &TO&-&\textbf{37.25}&\textbf{0.15} \\
                rb-floodMin\_V0&10&7&\cmark& & & & &\textbf{0.08}&\textbf{0.03}&\textbf{0.08}&\textbf{0.03} \\
                rb-floodMin\_V1&10&7&\cmark& & & & &\textbf{0.08}&\textbf{0.03}&\textbf{0.08}&\textbf{0.03} \\
                rb-RelBrd\_V1&7&4&\cmark& & & & &\textbf{0.07}&\textbf{0.03}&\textbf{0.07}&\textbf{0.03} \\
                rb-reset\_V0&47&26&\cmark& & & & &TO&-&\textbf{80.60}&\textbf{0.19} \\
                rb-reset\_V1&47&26&\cmark& & & & &TO&-&\textbf{236.44}&\textbf{0.22} \\
                rb-simple-2x\_reset\_V0&43&21&\cmark& & & & &11.52&0.32&\textbf{1.28}&\textbf{0.04} \\
                rb-simple-2x\_reset\_V1&43&21&\cmark& & & & &1138.11&31.25&\textbf{5.47}&\textbf{0.07} \\
                rb-simple-reset\_V0&39&19&\cmark& & & & &334.97&13.89&\textbf{2.27}&\textbf{0.04} \\
                rb-simple-reset\_V1&39&19&\cmark& & & & &239.67&13.16&\textbf{3.66}&\textbf{0.04} \\
                \hline
            \end{tabular}
        }
    \end{center}
\end{table}

\smartpar{Analysis: MTA Benchmarks}
On all benchmarks, except ``nbacg'' and ``nbacr'' from the \emph{Promela} set, at least one model checker of \toolname outperformed ByMC.

On the majority of the safe benchmarks ($15$ out of $19$), one or both of the ZCS and ACS model checkers were among the fastest.
This is mostly because for these TA the abstract error graphs constructed in those model checkers contained few paths that needed to be checked using the SMT solver or were even empty.

Notably, for the hand-coded ``cc'' benchmark, both the ACS and ZCS model checkers timed out.
For this benchmark, both error graphs contain many abstract error paths that have to be checked for spuriousness by ACS and ZCS.

Out of the nine unsafe benchmarks, four were solved fastest by the SMT model checker, two by ByMC, and two by the ZCS approach.
In most of these cases, the ACS- and ZCS-based approaches timed out, since the error graph contains many abstract error paths and takes a lot of time to check them.

The two benchmarks that were solved fastest by ByMC were not solved within the timeout by the SMT-based approach, but they were solved in less than a minute by the ZCS-based approach.
Here, the formula $\phi_{REACH}$ is very big, whereas the error graph and the number of abstract error paths in the ZCS approach remain feasible.

Surprisingly, ByMC reported unsupported syntax for the ``cc\_case*'' benchmarks from \emph{Promela}.
The cause was unclear, since the files were generated using ByMC, appear valid, and were accepted by \toolname.

In summary, these results show that on unsafe benchmarks, the SMT-based approach outperforms most model checkers, with some exceptions where the ZCS approach is faster.
On safe benchmarks, the ACS and ZCS approaches outperform SMT and ByMC, with ACS generally being more efficient on small benchmarks and ZCS better on larger ones.

\smartpar{Analysis: ETA Benchmarks}
The ETA benchmarks are only evaluated on the ZCS and ACS model checkers, as the other approaches do not support extended threshold automata.
The ZCS approach outperforms the ACS  model checker on all safe ETA benchmarks except for the three smallest benchmarks, where both tie.
Closer analysis indicates that the ACS approach takes significantly longer to construct the error graph on larger examples, suggesting that $01$-abstraction paired with the efficient representation as BDDs is more efficient than the WSTS approach for large benchmarks.

On unsafe benchmarks with limited error graph size, the ACS approach finds counterexamples faster.
This is most likely because the error graph exploration of the ACS model checker uses depth-first instead of breadth-first search, finding a concrete counterexample much quicker. 

In summary, for larger examples, the ZCS model checker outperforms the ACS model checker, whereas on small unsafe ones, the ACS model checker can be significantly faster.

%% file: small-conclusion.tex
\section{Conclusion}

We introduced \toolname, a modular, openly available, well-documented toolsuite for the verification of threshold-based distributed algorithms. Our implementation includes multiple model checkers for decidable fragments, as well as semi-decision procedures that extend beyond these fragments. In our experimental evaluation, \toolname significantly outperformed the only other existing but no longer maintained tool, ByMC.  

As future work, we plan to add support for liveness verification, including the development of novel algorithms that reduce the proof burden on protocol designers compared to the existing approaches.
On the user experience side, we are working on an automatic translation from pseudocode or domain-specific languages to threshold automata, as well as a graphical user interface.
These new interfaces should help to make \toolname more accessible for non-expert users.

%% file: example.tex
\section{Modeling Example: Reliable Broadcast Protocol}
\label{apx:rbp}
\input{resources/floodmin-alg-ta-new.tex}

\begin{example}
\cref{fig:alg-floodmin2} shows the pseudocode of a reliable broadcast, inspired by \cite{srikanth1987simulating}.
In every round, processes with input $v=1$ will broadcast a message, then processes with $v=0$ will set $v=1$ if they received at least $t+1$ messages, and finally processes set $accept=true$ if at least $n-t$ messages have been received. If $accept$ is false at the end of the round, a new round starts. 

\cref{fig:ta-floodmin2} depicts a TA for this algorithm, with 
$L = \{V_0, V_1, RV_0, SE, AC\}$, $\mI = \{V_0, V_1\}$, $\Shared = \{nsnt, rec\}$, $\Pi = \{n, t, f\}$, and 
$RC = n > 3t \land t \geq f \geq 0$.
A process in $V_1$ has input $1$ and can move freely (there is no guard) to $SE$, incrementing both variables $nsnt$ and $rec$ to simulate lines~4--6 of the algorithm. 
A process in location $V_0$ has input $0$ and can move to $RV_0$, incrementing $rec$ to simulate line~6 (but not line 5, since the condition in line~4 evaluates to false). 
From $RV_0$, a process can move to $SE$ if $nsnt \geq \varphi_1$, simulating lines~7-8. 
Processes that started with $v=1$ already are in $SE$, corresponding to the fact that $v=1$ is not changed in line 8.

Note that instead of being $t+1$, $\varphi_1$ is chosen as $t + 1 - f$. %
This prevents processes from making a transition based on messages from faulty processes, hence the TA represents the behavior of correct processes and $f$ the effect of faulty processes on correct ones.
For more details on the role of $f$, see \cite{Stoilkovska0WZ20}. 

Processes in $SE$ can move to $AC$ if $nsnt \geq \varphi_2$, simulating lines~9-10. 
If $nsnt < \varphi_2$ and $rec \geq \varphi_3$ (this constraint corresponds to waiting long enough for all non-faulty processes to receive all messages), the condition in line~9 was not satisfied and a new iteration of the while-loop is started by moving back from $SE$ to $V_1$.
The first process taking this transition will reset $nsnt$ and $rec$, the others take the transition that does not change any variables.
Similarly, processes in $RV_0$ will move to $V_0$ to start the new round.

The $RC$ condition is critical here to ensure fault-tolerance: if $f > t$, then $AC$ will be reachable even if all processes start in $V_0$, violating the validity property.
\end{example}

%% file: resources/floodmin-alg-ta-new.tex
 \begin{figure}[h]
  \centering
  \begin{minipage}[t]{0.49\textwidth}
  \vspace{0pt} 
    \centering
    \begin{algorithm}[H]
        \caption{}
        \begin{algorithmic}[1]
        \State \textbf{int} $v = \text{input}(\{0, 1\})$
        \State \textbf{bool} accept = false;
        \While{$(!accept)$} 
            \If{$(v == 1)$}
                \State broadcast $\langle \text{ECHO} \rangle$;
            \EndIf
            \State receive messages from all;
            \If{received $\langle \text{ECHO} \rangle$ $\geq t + 1$}
                \State $v = 1$;
            \EndIf
            \If{received $\langle \text{ECHO} \rangle$ $\geq n - t$}
                \State accept = true;
            \EndIf
        \EndWhile
        \end{algorithmic}
        \end{algorithm}
        \vspace{-25pt}
    \caption{Pseudocode of a reliable broadcast protocol.}
    \label{fig:alg-floodmin2}
  \end{minipage}
  \hfill
  \begin{minipage}[t]{0.49\textwidth}
    \vspace{21.75pt} 
    \scalebox{.80}{
        \begin{tikzpicture}[thick, lts, font=\scriptsize]
            \node[location, initial, initial text=] (V0) at (0,0) {$V_0$};
            \node[location, initial, initial text=] (V1) at (0,-4.5) {$V_1$};

            \node[location] (RV0) at (3.5,0) {$RV_0$};

            \node[location] (SE) at (3.5, -2.25) {$SE$};

            \node[location] (AC) at (6, -2.25) {$AC$};

            \path[->]
                (V0) edge [above,bend left=10] node [] {$rec{++}$} (RV0)
                (V1) edge [below, bend right=50, sloped] node [] {$nsnt{++};rec{++}$} (SE)
                (RV0) edge [right] node {$nsnt \geq \varphi_1$} (SE)
                (SE) edge [] node {$nsnt \geq \varphi_2$} (AC)
                (SE) edge [bend right=10,sloped] node [text width=30mm, yshift=-1mm]{$rec \geq \varphi_3  \wedge nsnt < \varphi_2$\\$nsnt := 0; rec := 0$} (V1)
                (SE) edge [bend left=10,sloped,below] node []{$rec = 0$} (V1)
                (RV0) edge [bend left =35, below] node [text width=30mm, xshift=1.5mm] {$rec \geq \varphi_3  \wedge nsnt < \varphi_1$\\$nsnt := 0; rec := 0$} (V0)
                (RV0) edge [bend left=12.5, below] node [yshift=0.5mm] {$rec = 0$} (V0)
            ;

            \node[align=center] (r) at (5.4, -4.75) {
                    \newcolumntype{a}{>{\columncolor{Gray}}r}
                    \newcolumntype{b}{>{\columncolor{Gray}}l}
                    \begin{tabular}{|ab|}
                        \hline
                        $\varphi_1$ & $= t - f + 1$ \\
                        $\varphi_2$ & $= n - t - f$ \\
                        $\varphi_3$ & $= n - f$ \\
                        \hline
                    \end{tabular}
            };
        \end{tikzpicture}
    }%
    \vspace{2.5pt}
    \caption{TA of the algorithm in \cref{fig:alg-floodmin}.}
    \label{fig:ta-floodmin2}
  \end{minipage}
\end{figure}

%% file: input-format.tex
\section{Input Formats}

This section presents the threshold automaton from \cref{fig:ta-floodmin} and \cref{fig:ta-floodmin2} into the two input formats supported by \toolname.

\subsection{\texttt{.ta} Input}\label{apx:eta}

\begin{lstlisting}[caption={TA of \cref{fig:ta-floodmin2} encoded as a \texttt{.ta} file},captionpos=b, label=lst:floddmin]
ta SRB {
    shared nsnt, rec;
    parameters n, t, f;
    assumptions (3) { n > 3 * t; t >= f; f >= 0; }
    locations (5) { V0: [0]; V1: [1]; RV0: [2]; SE: [3]; AC: [4]; }
    inits(6) {  nsnt == 0; rec == 0; RV0 == 0; SE == 0; AC == 0;
        V1 + V0 == n - f; /* n-f correct processes in initial states */ }
    rules (8) {
        0: V0 -> RV0 when (true) do { rec' := rec + 1; };
        1: V1 -> SE when (true) 
            do { nsnt' := nsnt + 1; rec' := rec + 1; };
        2: RV0 -> SE when (nsnt >= t + 1 - f) do {};
        3: SE -> AC when (nsnt >= n - t - f) do {};
        4: RV0 -> V0 when ((rec >= n - f) && (nsnt > t + 1 - f)) do {};
        5: RV0 -> V0 when (rec == 0) do {};
        6: SE -> V1 when ((rec >= n - f) && (nsnt > t + 1 - f)) 
            do { rec' := 0; nsnt' := 0; };
        7: SE -> V1 when (rec == 0) do {};
    }
    specifications (1) { validity:   V1 == 0 -> [](AC == 0); }
}
\end{lstlisting}

\subsection{TLA+ Input}\label{apx:tla}

\begin{lstlisting}[language=tla, caption={TLA$^+$ specification of \cref{fig:ta-floodmin}}]
---------------------------- MODULE ALG1---------------------------
EXTENDS Integers, FiniteSets
CONSTANT Processes, NbOfCorrProc, n, t

ASSUME NbOfCorrProc = n
    /\ n > 3 * t

VARIABLES ProcessesLocations, x0, x1

TypeOk == x0 \in Nat /\ x1 \in Nat
    /\ ProcessesLocations \in [Processes -> {"V0", "V1", "D0", "D1", "WAIT"}] 

Init == ProcessesLocations \in [Processes -> {"V0", "V1"}]
    /\ x0 = 0 /\ x1 = 0
------------------------------------------------------------------
Rule0(p) == ProcessesLocations[p] = "V0"
    /\ ProcessesLocations' = [ProcessesLocations EXCEPT ![p] = "WAIT"]
    /\ x0' = x0 + 1
    /\ UNCHANGED <<x1>>                         
------------------------------------------------------------------
Rule1(p) == ProcessesLocations[p] = "V1"
    /\ ProcessesLocations' = [ProcessesLocations EXCEPT ![p] = "WAIT"]
    /\ x1' = x1 + 1
    /\ UNCHANGED <<x0>>               
------------------------------------------------------------------
Rule2(p) == ProcessesLocations[p] = "WAIT"
    /\ x0 >= n - t          
    /\ ProcessesLocations' = [ProcessesLocations EXCEPT ![p] = "D0"]
    /\ UNCHANGED <<x0,x1>> 
------------------------------------------------------------------
Rule3(p) == ProcessesLocations[p] = "WAIT"
    /\ x1 >= n - t          
    /\ ProcessesLocations' = [ProcessesLocations EXCEPT ![p] = "D1"]
    /\ UNCHANGED <<x0,x1>> 
------------------------------------------------------------------
Next == \E p \in Processes: Rule0(p) \/ Rule1(p) \/ Rule2(p) \/ Rule3(p) 
                            
Spec == Init /\ [][Next]_<< ProcessesLocations,x0,x1 >>
------------------------------------------------------------------
NumInD0 == Cardinality({p \in Processes : ProcessesLocations[p] = "D0"})
NumInD1 == Cardinality({p \in Processes : ProcessesLocations[p] = "D1"})

cor == []( ~( NumInD0 > 0 /\ NumInD1 > 0))
\end{lstlisting}

\begin{lstlisting}[language=tla, caption={TLA$^+$ specification of \cref{fig:ta-floodmin2}}]
---------------------------- MODULE SRB ---------------------------
EXTENDS Integers, FiniteSets
CONSTANT Processes, NbOfCorrProc, N, T, F

ASSUME NbOfCorrProc = N - F
    /\ N > 3 * T
    /\ T >= F
    /\ F >= 0

VARIABLES ProcessesLocations, nsnt, rDone

TypeOk == rDone \in Nat /\ nsnt \in Nat
    /\ ProcessesLocations \in [Processes -> {"V0", "V1", "SE", "AC", "fRound"}] 

Init == ProcessesLocations \in [Processes -> {"V0"}]
    /\ nsnt = 0 /\ rDone = 0
------------------------------------------------------------------
Rule0(p) == ProcessesLocations[p] = "V0"
    /\ nsnt >= T - F + 1
    /\ ProcessesLocations' = [ProcessesLocations EXCEPT ![p] = "SE"]
    /\ nsnt' = nsnt + 1
    /\ UNCHANGED <<rDone>>                         
------------------------------------------------------------------
Rule1(p) == ProcessesLocations[p] = "V0"
    /\ nsnt >= N - T - F
    /\ ProcessesLocations' = [ProcessesLocations EXCEPT ![p] = "AC"]
    /\ UNCHANGED <<nsnt, rDone>>               
------------------------------------------------------------------
Rule2(p) == ProcessesLocations[p] = "V1"                  
    /\ ProcessesLocations' = [ProcessesLocations EXCEPT ![p] = "SE"]
    /\ nsnt' = nsnt + 1
    /\ UNCHANGED <<rDone>> 
------------------------------------------------------------------
Rule3(p) == ProcessesLocations[p] = "V1"
    /\ nsnt >= N - T - F
    /\ ProcessesLocations' = [ProcessesLocations EXCEPT ![p] = "AC"]
    /\ UNCHANGED <<nsnt, rDone>>
------------------------------------------------------------------
Rule4(p) == ProcessesLocations[p] = "SE"
    /\ nsnt >= N - T - F                    
    /\ ProcessesLocations' = [ProcessesLocations EXCEPT ![p] = "AC"]
    /\ UNCHANGED <<nsnt, rDone>>                   
------------------------------------------------------------------
Rule5(p) == ProcessesLocations[p] = "AC"
    /\ ProcessesLocations' = [ProcessesLocations EXCEPT ![p] = "fRound"]
    /\ rDone' = rDone + 1 
    /\ UNCHANGED <<nsnt>>        
------------------------------------------------------------------
Rule6(p) == ProcessesLocations[p] = "fRound"
    /\ rDone >= N - F
    /\ ProcessesLocations' = [ProcessesLocations EXCEPT ![p] = "V1"]
    /\ nsnt' = 0
    /\ rDone' = 0
------------------------------------------------------------------
Rule7(p) == ProcessesLocations[p] = "fRound"
    /\ rDone > 1
    /\ ProcessesLocations' = [ProcessesLocations EXCEPT ![p] = "V1"]
    /\ UNCHANGED <<nsnt, rDone>>                     
------------------------------------------------------------------
Next == \E p \in Processes: Rule0(p) \/ Rule1(p) \/ Rule2(p) \/ Rule3(p) 
    \/ Rule4(p) \/ Rule5(p) \/ Rule6(p) \/ Rule7(p)
                            
Spec == Init /\ [][Next]_<< ProcessesLocations,nsnt,rDone >>
------------------------------------------------------------------

NumInV1 == Cardinality({p \in Processes : ProcessesLocations[p] = "V1"})
NumInAC == Cardinality({p \in Processes : ProcessesLocations[p] = "AC"}) 

validity1 == 
        NumInV1 = 0 => [](NumInAC = 0)
\end{lstlisting}

%% file: extended-evaluation.tex
\section{Extended Evaluation}
\label{sec:extended-evaluation}

This section extends the evaluation in \cref{sec:benchmarks}. We provide more details on the setup, present additional results on an extended benchmark set and evaluate additional modes of \toolname and ByMC.

\subsection{Evaluation Environment}
\label{sec:evaluation-environment}

The evaluation was conducted on Dell R6525 nodes equipped with 2x AMD Epyc 7773x with 128 physical cores + 128 Simultaneous Multithreading and 2TB of RAM. The benchmarking scripts always report the elapsed real-time and maximal resident set size as reported by the GNU \texttt{time} command.
All model checkers were set to run in their sequential modes.

Additionally, we used the \texttt{timeout} command to stop benchmark runs that exceed the maximal runtime. 
During the evaluation, the timeout was set to $20$min, and a memory limit (on the virtual memory consumption) of $2071552$MB was set using \texttt{ulimit -SHv}.

\smartpar{ByMC Container}
For benchmarking ByMC, we created a container image of the tool from a custom Dockerfile\footnote{Available in the reproduction package \cite{tacoartifact} or on GitHub \url{https://github.com/pleich/bymc/blob/0bca349cc7d3e8511f649726be4ce990f458c00a/bymc/Dockerfile} (accessed 16-01-2026)}. 
Alternatively, a link to a virtual machine (VM) is provided in the ByMC repository\footnote{Link available in top level \texttt{README} section ``Installation'' in the ByMC GitHub repository~\cite{bymcgithub}}. 
We chose to create our own Dockerfile for two main reasons: 
\begin{itemize}
    \item Newer dependencies.
        To reduce the impact of optimizations to external components (like SMT solvers), we included the most recent version for which the build of ByMC would succeed without modification to the source code. 
        The VM comes, for example, with Z3~\cite{z3} \texttt{4.4.1} (released Oct 5th 2015 on Github\footnote{\url{https://github.com/Z3Prover/z3/releases/tag/z3-4.4.1}, accessed 16-01-2026}) installed, which is three years older than the version included in the container image, which is \texttt{4.7.1} (released May 23rd 2018 on GitHub\footnote{\url{https://github.com/Z3Prover/z3/releases/tag/z3-4.7.1}, accessed 16-01-2026}).
        Note that the Z3 version in the VM is surprising since the ByMC tool paper \cite{0001W18} explicitly reports that the benchmark results were obtained with Z3 version \texttt{4.6.0}.
        
        Still, ByMC has not been maintained for years, and many dependencies cannot be easily upgraded to up-to-date versions.
    \item Cumbersome benchmark process. 
        The ByMC VM is based on Debian 9 and in our testing, the guest additions did not work properly. 
        This made automation of the rather large set of benchmarks difficult.
\end{itemize}

Any errors encountered during evaluation, like, for example, the syntax errors reported in \cref{sec:benchmarks} and the errors described in the next section, were reproducible in the VM.

\subsection{ByMC Additional Modes}
\label{sec:exclusion-cav15}

ByMC, as provided in the VM and in the Dockerfile, has support for multiple modes.
The default one is \texttt{ltl} (implementing the methods described in \cite{KonnovLVW17,KonnovLVW16}), which we used for the evaluation in \cref{sec:benchmarks}.
Besides the \texttt{ltl} mode, ByMC implements a mode \texttt{ltl-mpi} (corresponding to \cite{0001W18}) and a mode \texttt{cav15} (corresponding to \cite{KonnovVW15,fmsdKonnovLVW17}).

The \texttt{cav15} mode is intended for verification of safety properties (specifically reachability as defined in \cite{KonnovVW15}) and should be compatible with our set of benchmarks.
However, as mentioned in \cref{sec:benchmarks}, the latter mode was excluded from the evaluation as it reports (invalid) counterexamples to safe benchmarks (which are also reported as safe by the \texttt{ltl} mode and \toolname).

\begin{figure}
    \centering
    \includegraphics[width=0.95\textwidth]{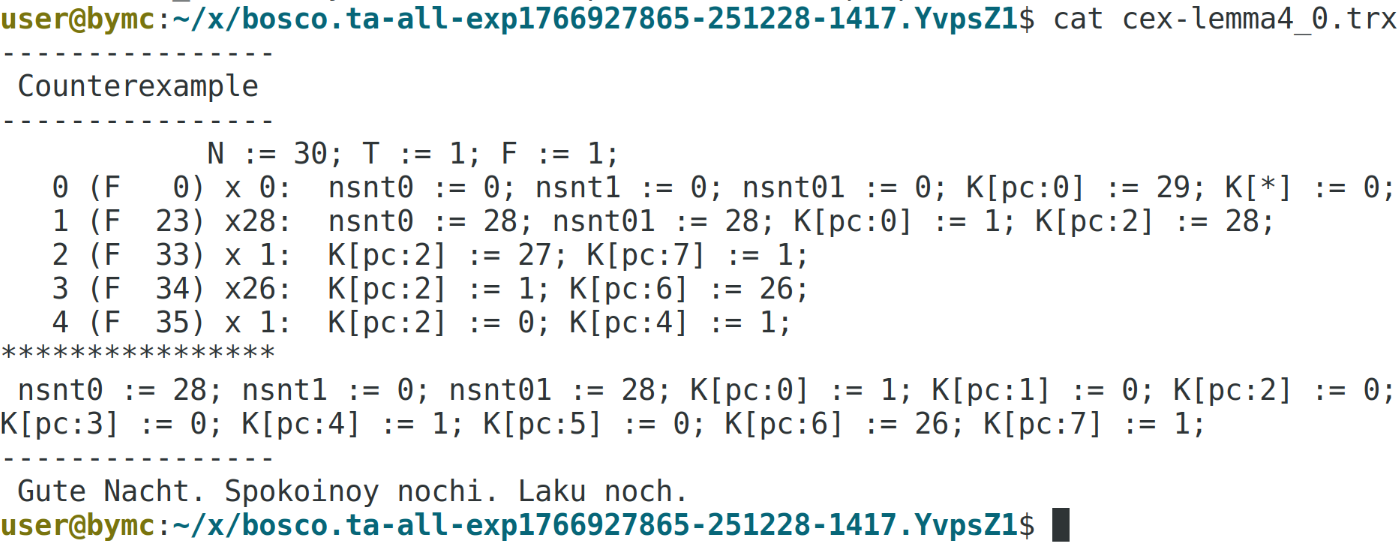}
    \caption{Screenshot of the counterexample reported ByMC in the \texttt{cav15} mode on \texttt{isola18/bosco.ta} (running in the VM provided by the ByMC authors\cite{bymcgithub})}
    \label{fig:bymc-spurious-cex}
\end{figure}

Specifically, this mode reports counterexamples for three hand-coded benchmarks: \texttt{bosco.ta}, \texttt{c1cs.ta} and \texttt{cf1s.ta} (from the \texttt{fault-tolerant-benchmarks} repository in folder \texttt{isola18/ta}\cite{benchmarksgithub}).
For example, for the \texttt{bosco.ta} ByMC reports the counterexample depicted in \cref{fig:bymc-spurious-cex}.
This counterexample does not correspond to a valid run in the threshold automaton.

Because of these correctness issues, we did not include this mode in the evaluation in \cref{sec:benchmarks}. 
However, we still include performance result in \cref{tab:extended-benchmark-ta}.

\smartpar{Older ByMC Modes} 
In addition to the modes mentioned above, at some point ByMC supported additional (older) decision procedures described in \cite{JohnKSVW13,KonnovVW14,KonnovVW17,KonnovVW15,fmsdKonnovLVW17,memKonnovVW15,0001LSW23} that used a refinement loop and counter abstraction combined with different model checkers. 
However, as mentioned in \cite{KonnovVW15,fmsdKonnovLVW17}, these older methods did not scale to larger examples.

Additionally, they are no longer executable inside the ByMC VM and we did not manage to compile the required dependencies of these modules (which are also marked as legacy in the ByMC repository~\cite{bymcgithub}).
Therefore, we do not report performance results for these modes.

\subsection{Origin of the Benchmarks}
\label{sec:origin-benchmarks}

All ``ByMC'' benchmarks appearing in \cref{tab:extended-benchmark-ta} were directly taken or generated from the \texttt{fault-tolerant-benchmarks} GitHub repository~\cite{benchmarksgithub}. 
These benchmarks originally appeared in \cite{JohnKSVW13,KonnovVW14,KonnovVW17,KonnovVW15,fmsdKonnovLVW17,KonnovLVW17,KonnovLVW16,0001W18,BertrandKLW19}. 

For benchmarks in the Promela format, ByMC was used to first create the threshold automata in the \texttt{.ta} format (via data abstraction \cite{JohnKSVW13,KonnovVW14,KonnovVW17,KonnovVW15,fmsdKonnovLVW17}).
The benchmarking was then conducted directly on the \texttt{.ta} files.
When benchmarking ByMC, all liveness properties were removed from the files.
The \texttt{.ta} files used during the evaluation are included in the reproduction package~\cite{tacoartifact} (and in the TACO GitHub repository \footnote{\url{https://github.com/cispa/TACO}}).

The ``RedBelly'' and ``King Consensus'' benchmarks first appeared in \cite{BaumeisterEJSV24} and are also available in the TACO GitHub repository and in the reproduction package~\cite{tacoartifact}.

\smartpar{Selection of Benchmarks in \cref{sec:benchmarks}}
Note that many benchmarks in this full set of benchmarks are quite similar. 
For example, the \texttt{README} for the benchmarks \emph{ByMC POPL17 Promela} states that they are an extension of the benchmark set \emph{ByMC CAV15 Promela} with some corrections to correct for liveness properties. This made it difficult to select a representative set.

Threfore, for the ByMC benchmarks in \cref{sec:benchmarks}, we chose to use the same benchmark set as in \cite{0001W18} and \cite{Balasubramanian20}.
We believe that the files named ``cc'' correspond to the benchmarks called ``cbc'' in \cite{0001W18}, and we could not obtain the two additional ``bosco'' cases mentioned there. Note that they are also missing from \cite{Balasubramanian20}.

\input{resources/extended-evaluation-results}

%% file: resources/extended-evaluation-results.tex
\subsection{Extended Evaluation Results}
\label{sec:extended-eval-results}

The execution times  and memory consumption of ByMC in the default (\texttt{ltl}) and in the \texttt{cav15} mode, as well as the execution times for all of \toolname's model checkers with Z3~\cite{z3} and with CVC5~\cite{cvc5} as SMT solver are provided in \cref{tab:extended-benchmark-ta}.

\smartpar{Analysis of MTA Benchmarks \cref{tab:extended-benchmark-ta}} Out of the 86 benchmarks, one of \toolname's model checkers outperformed ByMC (in either mode) in $59$ cases, $64$ if we exclude the \texttt{cav15} mode. 
Moreover, there is no benchmark that can be solved by ByMC but is not solved by any model checker of \toolname within the 20min timeout. 
In contrast, there are eight benchmarks (not counting the cases where ByMC returned errors), which none of the modes of ByMC solved within the time limit.

Otherwise, the extended evaluation confirms the observations in \cref{sec:benchmarks}. For unsafe benchmarks, usually the SMT model checker performs best, with some exceptions where the ZCS model checker is faster. On safe benchmarks, ACS and ZCS are often faster than the SMT model checker.

\smartpar{Random19 Benchmarks} Note that most of the benchmarks where ByMC outperforms \toolname stem from the \emph{ByMC Random19\cite{BertrandKLW19}} set of benchmarks. These benchmarks have two properties that make them particularly challenging for our model checkers.

On the one hand, the SMT model checker does not perform well because these benchmarks have a high number of potential context switches that could occur on an error path (denoted as the number $k$ in \cite{Balasubramanian20}). Every additional context switch adds variables to the SMT query and increases its length. 

On the other hand, these examples also have many potential abstract interval orderings. As the ZCS and ACS model checker have to check reachability in every interval order, resulting in many individual model checker runs (for more details, refer to \cite{BaumeisterEJSV24}).
Most likely the performance for these benchmarks could be further improved by developing a preprocessor tailored towards these examples.

\smartpar{CVC5 vs. Z3} 
When comparing the performance of the SMT model checker when paired with CVC5~\cite{cvc5} and Z3~\cite{z3}, we observe that on larger examples, like \emph{ByMC CAV15 Promela} or \emph{ByMC ISOLA18 Promela}, the model checker paired with CVC5 mostly outperforms Z3 while usually consuming less memory. 
CVC5 seems to scale better to the large SMT queries generated by the SMT model checker for these examples. 
Consequently, our experiments suggest that the SMT model checker should be paired with CVC5 when verifying large TA.

On smaller benchmarks, like the benchmarks in the \emph{ByMC Handcoded} \newline \emph{ISOLA18} set, Z3 is in most cases faster than CVC5. For the ``bosco'' case in particular, the SMT model checker paired with Z3 is around four times faster. 

Next, we compare the runtime of the ACS and ZCS model checker when paired with different SMT solvers.  It is important to note that the SMT queries can significantly differ from the queries of the SMT model checker. Generally, there will be more queries but with significantly fewer variables and clauses. 
This is becuase the abstract error graphs already provide information on which rules can be applied when checking for spurious counterexamples. However, there might be many paths that need to be checked.
Additionally, both ACS and ZCS rely on constructing all potential orders between abstract interval borders. This step additionally requires many small SMT queries. 

Here, Z3 almost always outperforms the same model checker when paired with CVC5. Most likely because it is optimized to solve small queries fast. Therefore, it is generally advisable to select Z3 for these model checkers.

\smartpar{Examples not Supported by ACS} \cref{tab:extended-benchmark-ta} contains two examples, ``toy'' and ``nbacg-asyn-guer01-nbac'', which are not supported by the ACS model checker. 
These examples fullfill the conditions mentioned in \cref{sec:implementation}, i.e., they contain properties that require some locations in a target configuration to be empty.
However, the wqo used in the ACS model checker is not precise for locations for these cases \cite{BaumeisterEJSV24}. 
This could potentially be addressed in the future by utilizing a different wqo, but would require additional theory. Note that such a wqo will also result in larger error graphs.
As properties containing such constraints rarely occur, the ACS model checker has been implemented with the wqo from \cite{BaumeisterEJSV24}.

The analysis of the benchmark results for the ETA benchmarks can be found in \cref{sec:benchmarks} as the complete list of ETA benchmarks has already been included in \cref{tab:benchmark-ta}.

\begin{landscape}
    \begin{table}[!ht]
        \vspace*{-27mm}
        \caption{Extended benchmark results for \toolname and ByMC. Columns $|L|$ and $|\Rules|$ give the number of locations and rules of the given TA. In column ``safe?'', \cmark\xspace denotes that all checked properties hold, \xmark\xspace denotes that some property was violated, and $?$ that the result is unknown. Execution time (here elapsed wall clock time) is given in s, \emph{TO} denotes a timeout after 20min. The fastest execution time per benchmark is highlighted in bold font. RSS columns give the maximal resident set size in GB. ``Err'' denotes that an error occurred, and ``Unsup.'' that the benchmark is not supported. Entries in \textcolor{red}{red} are benchmarks of ByMC in \texttt{cav15} mode which reported invalid counterexamples (see \cref{sec:exclusion-cav15}).}
        \label{tab:extended-benchmark-ta}
        \centering
        \scalebox{0.8}{
            \begin{tabular}{|l | g c g || r g | r g || r g | r g | r g | r g | r g | r g|}
                \hline
                \multirow{4}{*}{\textbf{Benchmark}} & \cellcolor{white} & & \cellcolor{white} & \multicolumn{4}{c||}{\multirow{2}{*}{\textbf{ByMC}}}& \multicolumn{12}{c|}{\textbf{TACO}} \\
                \cline{9-20}
                    &\cellcolor{white} & & \cellcolor{white} & \multicolumn{4}{c||}{}  & \multicolumn{4}{c|}{\textbf{SMT}} & \multicolumn{4}{c|}{\textbf{ACS}} & \multicolumn{4}{c|}{\textbf{ZCS}} \\
                \cline{5-20}
                    & \multicolumn{3}{c||}{\multirow{-2}{*}{\textbf{TA}}}  & \multicolumn{2}{c|}{\textbf{\texttt{ltl}}} & \multicolumn{2}{c||}{\textbf{\texttt{cav15}}} & \multicolumn{2}{c|}{\textbf{CVC5}} & \multicolumn{2}{c|}{\textbf{Z3}} & \multicolumn{2}{c|}{\textbf{CVC5}} & \multicolumn{2}{c|}{\textbf{Z3}} & \multicolumn{2}{c|}{\textbf{CVC5}} & \multicolumn{2}{c|}{\textbf{Z3}} \\ 
                    & $|L|$ & $|\Rules|$ & safe? & time & RSS  & time & RSS  & time & RSS & time & RSS & time & RSS & time & RSS & time & RSS & time & RSS \\
                \hline\hline
                \textbf{MTA Benchmarks}& & & & & & & & & & & & & & & & & & & \\
                \hline
                \textbf{ByMC CONCUR14 Promela\cite{KonnovVW14,KonnovVW17}} & & & & & & & & & & & & & & & & & & & \\
                    asyn-byzagreement0&37&202&\cmark&Err&-&Err&-&0\textbf{0.77}&\textbf{0.10}&2.27&0.27&1.84&0.05&1.68&0.03&1.67&0.05&1.40&0.04 \\
                    asyn-ray97-nbac-clean&109&1831&\xmark&\textbf{5.23}&\textbf{0.43}&6.75&0.42&TO&-&TO&-&TO&-&TO&-&30.55&0.25&21.25&0.30 \\
                    asyn-ray97-nbac&77&1431&\xmark&\textbf{3.09}&\textbf{0.25}&4.11&0.25&TO&-&TO&-&TO&-&TO&-&20.36&0.27&15.70&0.27 \\
                    bcast-byz&7&21&\cmark&0.34&0.04&0.31&0.03&0.13&0.05&0.08&0.03&0.11&0.05&\textbf{0.06}&\textbf{0.03}&0.12&0.05&0.07&0.03 \\
                    bcast-folklore-crash&6&15&\cmark&0.31&0.04&0.29&0.03&0.10&0.05&0.07&0.03&0.10&0.05&\textbf{0.05}&\textbf{0.03}&0.10&0.05&0.06&0.03 \\
                    cond-consensus2&115&991&?&Err&-&Err&-&TO&-&TO&-&TO&-&TO&-&TO&-&TO&- \\
                    toy&5&8&\cmark&0.28&0.04&0.29&0.03&0.12&0.05&\textbf{0.07}&\textbf{0.03}&Unsup.&-&Unsup.&-&0.36&0.05&0.22&0.05 \\
                \hline
                \textbf{ByMC CAV15 Promela\cite{KonnovVW15,fmsdKonnovLVW17}} & & & & & & & & & & & & & & & & & & & \\
                    aba-asyn-byzagreement0\_case1&37&202&\cmark&15.23&0.06&3.77&0.07&\textbf{0.73}&\textbf{0.10}&4.65&0.47&1.78&0.05&2.12&0.03&1.34&0.05&1.53&0.04 \\
                    aba-asyn-byzagreement0\_case2&61&425&\cmark&289.05&0.11&61.89&0.14&\textbf{3.51}&\textbf{0.29}&29.26&1.84&21.37&0.08&21.40&0.09&17.29&0.14&17.49&0.15 \\
                    bosco\_case1&58&394&\xmark&228.10&0.10&\textbf{25.92}&\textbf{0.15}&75.17&0.22&69.57&0.91&TO&-&TO&-&TO&-&TO&- \\
                    bosco\_case2&88&752&\xmark&TO&-&\textbf{488.24}&\textbf{0.50}&988.40&0.82&TO&-&TO&-&TO&-&TO&-&TO&- \\
                    bosco\_case3&62&442&\xmark&155.05&0.11&\textbf{36.17}&\textbf{0.15}&181.53&0.31&61.92&0.46&TO&-&TO&-&TO&-&TO&- \\
                    c1cs\_case1&125&2050&\xmark&TO&-&991.92&0.87&\textbf{428.99}&\textbf{8.42}&TO&-&TO&-&TO&-&945.44&0.33&820.44&0.30 \\
                    c1cs\_case2&84&964&\xmark&TO&-&180.06&0.40&\textbf{57.75}&\textbf{1.44}&TO&-&111.97&0.86&261.23&0.72&75.75&0.19&71.97&0.21 \\
                    c1cs\_case3&129&2194&\xmark&TO&-&TO&-&\textbf{1006.74}&\textbf{19.72}&TO&-&TO&-&TO&-&TO&-&TO&- \\
                    cbc-cond-consensus2\_case1&74&425&?&Err&-&Err&-&TO&-&TO&-&TO&-&TO&-&TO&-&TO&- \\
                    cbc-cond-consensus2\_case2&40&169&\xmark&Err&-&Err&-&64.73&0.58&\textbf{25.61}&\textbf{0.38}&TO&-&TO&-&TO&-&TO&- \\
                    cbc-cond-consensus2\_case3&115&991&?&Err&-&Err&-&TO&-&TO&-&TO&-&TO&-&TO&-&TO&- \\
                    cbc-cond-consensus2\_case4&71&466&?&Err&-&Err&-&TO&-&TO&-&TO&-&TO&-&TO&-&TO&- \\
                    cf1s-consensus-folklore-onestep\_case1&57&438&\cmark&29.64&0.09&2.73&0.08&5.04&0.19&14.37&0.93&0.98&0.05&\textbf{0.89}&\textbf{0.03}&1.19&0.05&1.14&0.04 \\
                    cf1s-consensus-folklore-onestep\_case2&57&438&\cmark&172.04&0.14&7.43&0.11&5.33&0.23&13.38&0.50&1.26&0.05&\textbf{1.21}&\textbf{0.05}&4.02&0.05&3.65&0.04 \\
                    cf1s-consensus-folklore-onestep\_case3&98&1194&\cmark&TO&-&246.43&0.48&81.50&2.18&TO&-&33.33&0.59&29.36&0.59&\textbf{26.23}&\textbf{0.15}&47.45&0.20 \\
                    frb-bcast-folklore-crash&6&13&\cmark&0.36&0.04&0.29&0.03&0.11&0.05&0.07&0.03&0.11&0.05&\textbf{0.06}&\textbf{0.03}&0.11&0.05&\textbf{0.06}&\textbf{0.03} \\
                    nbac-asyn-ray97-nbac&77&1431&\xmark&\textbf{3.10}&\textbf{0.25}&4.08&0.26&TO&-&TO&-&TO&-&TO&-&21.26&0.27&19.16&0.27 \\
                    nbacc-asyn-ray97-nbac-clean&109&1831&\xmark&\textbf{5.15}&\textbf{0.43}&6.77&0.42&TO&-&TO&-&TO&-&TO&-&65.94&0.30&32.41&0.30 \\
                    nbacg-asyn-guer01-nbac&24&64&\xmark&\textbf{0.31}&\textbf{0.04}&0.32&0.03&0.57&0.06&0.43&0.06&Unsup.&-&Unsup.&-&0.76&0.05&0.56&0.05 \\
                    strb-bcast-byz&7&21&\cmark&0.35&0.04&0.31&0.03&0.12&0.05&\textbf{0.05}&\textbf{0.03}&0.12&0.05&0.06&0.03&0.12&0.05&0.07&0.03 \\
                \hline
                \textbf{ByMC POPL17 Promela\cite{KonnovLVW17,KonnovLVW16}} & & & & & & & & & & & & & & & & & & & \\
                    asyn-byzagreement0&37&202&\cmark&15.21&0.06&3.75&0.06&\textbf{0.76}&\textbf{0.10}&2.15&0.18&1.60&0.05&1.07&0.03&1.56&0.05&1.48&0.04 \\
                    asyn-guer01-nbac&24&64&\xmark&\textbf{0.33}&\textbf{0.04}&0.33&0.03&0.55&0.06&0.36&0.06&0.78&0.06&0.40&0.05&0.81&0.05&0.42&0.05  \\
                    asyn-ray97-nbac-clean&78&1431&\xmark&191.96&3.51&\textbf{3.82}&\textbf{0.36}&TO&-&TO&-&TO&-&TO&-&18.15&0.18&22.53&0.22 \\
                    asyn-ray97-nbac&77&1031&\xmark&\textbf{1.71}&\textbf{0.17}&2.23&0.16&TO&-&TO&-&TO&-&TO&-&12.00&0.18&8.12&0.22 \\
                    bcast-byz&7&21&\cmark&0.33&0.04&0.30&0.03&0.13&0.05&\textbf{0.07}&\textbf{0.03}&0.12&0.05&\textbf{0.07}&\textbf{0.03}&0.11&0.05&0.07&0.03 \\
                    bosco&28&152&\xmark&44.94&0.06&7.73&0.05&\textbf{5.58}&\textbf{0.11}&7.85&0.09&TO&-&495.32&0.07&125.78&0.19&104.55&0.17 \\
                    c1cs&101&1285&\xmark&TO&-&503.44&0.47&82.32&2.02&TO&-&TO&-&TO&-&\textbf{78.34}&\textbf{0.22}&103.71&0.22 \\
                    cond-consensus2&164&2064&?&Err&-&Err&-&TO&-&TO&-&TO&-&TO&-&TO&-&TO&- \\
                    consensus-folklore-onestep&41&280&\cmark&18.26&0.07&1.68&0.06&2.34&0.12&8.00&0.90&0.54&0.05&\textbf{0.46}&\textbf{0.03}&0.61&0.05&0.70&0.03 \\
                \hline
            \end{tabular}
    }
    \end{table}
\end{landscape}

\begin{landscape} 
    \hspace*{-15mm}
    \centering
    \scalebox{0.8}{
        \begin{tabular}{|l | g c g || r g | r g || r g | r g | r g | r g | r g | r g|}
            \hline
            \multirow{4}{*}{\textbf{Benchmark}} & \cellcolor{white} & & \cellcolor{white} & \multicolumn{4}{c||}{\multirow{2}{*}{\textbf{ByMC}}}& \multicolumn{12}{c|}{\textbf{TACO}} \\
            \cline{9-20}
                &\cellcolor{white} & & \cellcolor{white} & \multicolumn{4}{c||}{}  & \multicolumn{4}{c|}{\textbf{SMT}} & \multicolumn{4}{c|}{\textbf{ACS}} & \multicolumn{4}{c|}{\textbf{ZCS}} \\
            \cline{5-20}
                & \multicolumn{3}{c||}{\multirow{-2}{*}{\textbf{TA}}}  & \multicolumn{2}{c|}{\textbf{\texttt{ltl}}} & \multicolumn{2}{c||}{\textbf{\texttt{cav15}}} & \multicolumn{2}{c|}{\textbf{CVC5}} & \multicolumn{2}{c|}{\textbf{Z3}} & \multicolumn{2}{c|}{\textbf{CVC5}} & \multicolumn{2}{c|}{\textbf{Z3}} & \multicolumn{2}{c|}{\textbf{CVC5}} & \multicolumn{2}{c|}{\textbf{Z3}} \\ 
                & $|L|$ & $|\Rules|$ & safe? & time & RSS  & time & RSS  & time & RSS & time & RSS & time & RSS & time & RSS & time & RSS & time & RSS \\
            \hline\hline
            \textbf{ByMC Handcoded ISOLA18\cite{0001W18}} & & & & & & & & & & & & & & & & & & & \\
                aba&5&10&\cmark&0.69&0.04&0.33&0.03&0.13&0.05&0.07&0.03&0.12&0.05&\textbf{0.06}&\textbf{0.03}&0.12&0.05&0.07&0.03 \\
                bcrb&5&13&\cmark&0.42&0.04&0.29&0.03&0.12&0.05&0.08&0.03&0.11&0.05&\textbf{0.07}&\textbf{0.03}&0.12&0.05&\textbf{0.07}&\textbf{0.03} \\
                bosco&8&20&\cmark&139.33&0.07&\textcolor{red}{18.22}&\textcolor{red}{0.04}&4.39&0.06&\textbf{1.03}&\textbf{0.04}&TO&-&117.29&0.04&TO&-&160.34&0.05 \\
                c1cs&9&30&\cmark&TO&-&\textcolor{red}{44.72}&\textcolor{red}{0.23}&1.72&0.07&\textbf{0.72}&\textbf{0.05}&6.15&0.11&1.68&0.03&53.62&0.05&29.25&0.05 \\
                cc&7&14&\cmark&0.40&0.04&0.30&0.03&0.54&0.06&\textbf{0.26}&\textbf{0.04}&TO&-&TO&-&TO&-&TO&- \\
                cf1s&9&26&\cmark&397.75&0.10&\textcolor{red}{0.44}&\textcolor{red}{0.04}&0.33&0.06&0.22&0.04&0.32&0.05&\textbf{0.17}&\textbf{0.03}&0.51&0.05&0.28&0.05 \\
                nbacg&8&16&\cmark&0.36&0.04&0.29&0.03&0.41&0.05&\textbf{0.24}&\textbf{0.05}&0.51&0.05&0.24&0.05&0.70&0.05&0.41&0.05  \\
                nbacr&7&16&\cmark&0.31&0.04&0.27&0.03&0.11&0.05&\textbf{0.08}&\textbf{0.05}&0.13&0.05&\textbf{0.08}&\textbf{0.05}&0.16&0.05&0.11&0.05 \\
                strb&4&8&\cmark&0.26&0.04&0.29&0.03&0.10&0.05&0.06&0.03&0.10&0.05&0.06&0.03&0.10&0.05&\textbf{0.04}&\textbf{0.03} \\
            \hline
            \textbf{ByMC ISOLA18 Promela\cite{0001W18}} & & & & & & & & & & & & & & & & & & & \\
                aba\_case1&37&202&\cmark&15.19&0.06&3.75&0.07&\textbf{0.94}&\textbf{0.11}&2.08&0.17&1.72&0.05&1.19&0.03&1.47&0.05&1.43&0.04 \\
                aba\_case2&61&425&\cmark&290.42&0.11&63.37&0.14&\textbf{3.65}&\textbf{0.29}&19.49&0.74&14.95&0.08&17.79&0.09&15.30&0.15&15.12&0.15 \\
                bosco\_case1&28&152&\xmark&45.16&0.06&7.87&0.05&\textbf{5.74}&\textbf{0.10}&7.61&0.09&TO&-&332.93&0.06&124.56&0.15&104.76&0.19 \\
                bosco\_case2&40&242&\xmark&752.65&0.08&98.78&0.09&52.86&0.24&\textbf{38.47}&\textbf{0.21}&TO&-&TO&-&1178.32&0.30&TO&- \\
                bosco\_case3&32&188&\xmark&52.14&0.06&9.98&0.06&\textbf{7.24}&\textbf{0.11}&13.30&0.12&TO&-&982.39&0.10&156.98&0.17&117.45&0.17 \\
                c1cs\_case1&101&1285&\xmark&Err&-&Err&-&85.32&2.02&TO&-&TO&-&TO&-&\textbf{79.32}&\textbf{0.21}&83.00&0.22 \\
                c1cs\_case2&70&650&\xmark&TO&-&100.39&0.23&17.28&0.52&1164.87&14.06&TO&-&102.70&0.41&27.09&0.15&\textbf{26.23}&\textbf{0.15} \\
                c1cs\_case3&101&1333&\xmark&TO&-&TO&-&\textbf{156.59}&\textbf{3.95}&TO&-&TO&-&TO&-&TO&-&TO&- \\
                cc\_case1&164&2064&?&Err&-&Err&-&TO&-&TO&-&TO&-&TO&-&TO&-&TO&- \\
                cc\_case2&73&470&\xmark&Err&-&Err&-&TO&-&\textbf{1178.05}&\textbf{8.82}&TO&-&TO&-&TO&-&TO&- \\
                cc\_case3&304&6928&?&Err&-&Err&-&TO&-&TO&-&TO&-&TO&-&TO&-&TO&- \\
                cc\_case4&161&2105&?&Err&-&Err&-&TO&-&TO&-&TO&-&TO&-&TO&-&TO&- \\
                cf1s\_case1&41&280&\cmark&18.31&0.07&1.66&0.06&2.42&0.12&8.22&0.90&0.54&0.05&\textbf{0.43}&\textbf{0.03}&0.59&0.05&0.59&0.03 \\
                cf1s\_case2&41&280&\cmark&102.08&0.10&4.43&0.08&2.44&0.13&7.25&0.90&0.66&0.05&\textbf{0.54}&\textbf{0.03}&1.65&0.05&0.90&0.03 \\
                cf1s\_case3&68&696&\cmark&TO&-&108.28&0.20&22.75&0.71&41.35&1.17&\textbf{7.82}&\textbf{0.28}&7.99&0.28&10.79&0.09&8.53&0.09 \\
                frb&7&14&\cmark&0.36&0.03&0.32&0.03&0.11&0.05&\textbf{0.06}&\textbf{0.03}&0.11&0.05&0.06&0.03&0.11&0.05&0.06&0.03 \\
                nbacg&24&64&\xmark&0.30&0.04&\textbf{0.29}&\textbf{0.03}&0.55&0.06&0.40&0.06&0.78&0.06&0.41&0.05&0.91&0.05&0.41&0.05 \\
                nbacr&77&1031&\xmark&\textbf{1.69}&\textbf{0.17}&2.19&0.16&TO&-&TO&-&TO&-&TO&-&11.24&0.18&8.65&0.22 \\
                strb&7&21&\cmark&0.34&0.04&0.31&0.03&0.13&0.05&\textbf{0.07}&\textbf{0.03}&0.12&0.05&\textbf{0.07}&\textbf{0.03}&0.11&0.05&\textbf{0.07}&\textbf{0.03} \\
            \hline
        \end{tabular}
    }

    \newpage
    \centering
    \vspace*{-20mm}
    \scalebox{0.8}{
        \begin{tabular}{|l | g c g || r g | r g || r g | r g | r g | r g | r g | r g|}
            \hline
            \multirow{4}{*}{\textbf{Benchmark}} & \cellcolor{white} & & \cellcolor{white} & \multicolumn{4}{c||}{\multirow{2}{*}{\textbf{ByMC}}}& \multicolumn{12}{c|}{\textbf{TACO}} \\
            \cline{9-20}
                &\cellcolor{white} & & \cellcolor{white} & \multicolumn{4}{c||}{}  & \multicolumn{4}{c|}{\textbf{SMT}} & \multicolumn{4}{c|}{\textbf{ACS}} & \multicolumn{4}{c|}{\textbf{ZCS}} \\
            \cline{5-20}
                & \multicolumn{3}{c||}{\multirow{-2}{*}{\textbf{TA}}}  & \multicolumn{2}{c|}{\textbf{\texttt{ltl}}} & \multicolumn{2}{c||}{\textbf{\texttt{cav15}}} & \multicolumn{2}{c|}{\textbf{CVC5}} & \multicolumn{2}{c|}{\textbf{Z3}} & \multicolumn{2}{c|}{\textbf{CVC5}} & \multicolumn{2}{c|}{\textbf{Z3}} & \multicolumn{2}{c|}{\textbf{CVC5}} & \multicolumn{2}{c|}{\textbf{Z3}} \\ 
                & $|L|$ & $|\Rules|$ & safe? & time & RSS  & time & RSS  & time & RSS & time & RSS & time & RSS & time & RSS & time & RSS & time & RSS \\
            \hline\hline
            \textbf{ByMC Random19\cite{BertrandKLW19}} & & & & & & & & & & & & & & & & & & & \\
                ben-or&10&25&\cmark&0.65&0.04&\textbf{0.30}&\textbf{0.03}&4.56&0.08&1.93&0.07&TO&-&TO&-&TO&-&TO&- \\
                n-ben-or-byz&9&18&\cmark&0.43&0.04&\textbf{0.33}&\textbf{0.03}&5.98&0.08&1.59&0.05&TO&-&TO&-&TO&-&TO&- \\
                n-ben-or-nonclean&10&32&\cmark&0.93&0.05&\textbf{0.33}&\textbf{0.04}&12.61&0.10&3.60&0.07&TO&-&TO&-&TO&-&TO&- \\
                n-ben-or&10&27&\cmark&0.83&0.05&\textbf{0.33}&\textbf{0.03}&6.31&0.10&3.62&0.10&TO&-&TO&-&TO&-&TO&- \\
                n-kset&13&58&\cmark&39.36&0.10&\textbf{0.84}&\textbf{0.05}&68.77&0.26&32.10&0.23&TO&-&TO&-&14.72&0.12&16.59&0.12 \\
                n-rabc-cr&11&31&\cmark&1.01&0.05&\textbf{0.31}&\textbf{0.03}&61.13&0.15&9.65&0.13&TO&-&TO&-&TO&-&TO&- \\
                n-rabc-s&10&21&\cmark&0.40&0.04&\textbf{0.29}&\textbf{0.03}&17.79&0.11&2.71&0.07&TO&-&TO&-&TO&-&TO&- \\
                n-rabc&14&28&\xmark&TO&-&TO&-&49.34&0.18&\textbf{1.18}&\textbf{0.10}&TO&-&TO&-&TO&-&TO&- \\
                n-rs-bosco&19&48&\cmark&TO&-&TO&-&39.95&0.11&\textbf{9.00}&\textbf{0.09}&TO&-&TO&-&TO&-&TO&- \\
                p-ben-or-byz&9&16&\cmark&0.40&0.04&\textbf{0.36}&\textbf{0.03}&4.93&0.08&1.31&0.05&TO&-&TO&-&TO&-&TO&- \\
                p-ben-or-nonclean&10&30&\cmark&0.80&0.05&\textbf{0.32}&\textbf{0.04}&10.58&0.10&3.21&0.07&TO&-&TO&-&TO&-&TO&- \\
                p-ben-or&10&25&\cmark&0.71&0.05&\textbf{0.31}&\textbf{0.03}&4.77&0.08&2.44&0.07&TO&-&TO&-&TO&-&TO&- \\
                p-kset&13&52&\cmark&30.60&0.09&\textbf{0.79}&\textbf{0.05}&35.19&0.25&21.48&0.23&TO&-&TO&-&12.42&0.12&11.27&0.12 \\
                p-rabc-cr&11&29&\cmark&0.94&0.05&\textbf{0.32}&\textbf{0.03}&51.76&0.20&5.79&0.10&TO&-&TO&-&TO&-&TO&- \\
                p-rabc-s&10&19&\cmark&0.35&0.04&\textbf{0.31}&\textbf{0.03}&14.64&0.11&1.96&0.06&TO&-&TO&-&TO&-&TO&- \\
                p-rabc&14&28&\xmark&TO&-&TO&-&24.62&0.17&\textbf{1.48}&\textbf{0.10}&TO&-&TO&-&TO&-&TO&- \\
                p-rs-bosco&19&42&\cmark&TO&-&TO&-&32.49&0.10&\textbf{6.31}&\textbf{0.08}&TO&-&TO&-&TO&-&TO&- \\
            \hline
            \textbf{ByMC LMCS20\cite{0001LSW23}} & & & & & & & & & & & & & & & & & & & \\
                tendermint-1round-safe&6&22&\xmark&0.74&0.05&\textbf{0.33}&\textbf{0.03}&1.89&0.06&0.75&0.06&TO&-&891.86&0.19&TO&-&TO&- \\
            \hline
            \textbf{RedBelly small\cite{BaumeisterEJSV24}} & & & & & & & & & & & & & & & & & & & \\
                rb-bc&10&19&\cmark&0.96&0.04&TO&-&0.22&0.05&0.14&0.03&0.19&0.05&\textbf{0.09}&\textbf{0.03}&0.19&0.05&0.11&0.03\\
                rb-simple&19&33&\cmark&TO&-&TO&-&1.38&0.09&1.24&0.07&7.16&0.84&7.18&0.82&0.90&0.05&\textbf{0.81}&\textbf{0.03}\\
                rb&26&41&\cmark&TO&-&TO&-&6.29&0.16&10.04&0.49&63.96&3.45&58.47&3.44&1.16&0.05&\textbf{0.85}&\textbf{0.03}\\
            \hline\hline
                \textbf{ETA Benchmarks}& & & & & & & & & & & & & & & & & & & \\
            \hline
            \textbf{King Consensus\cite{BaumeisterEJSV24}} & & & & & & & & & & & & & & & & & & & \\
                phase-king-buggy&27&10&\xmark&&&&&&&&&52.19&3.31&\textbf{48.84}&\textbf{3.23}&TO&-&TO&- \\
                phase-king&27&10&\xmark&&&&&&&&&TO&-&\textbf{64.21}&\textbf{3.73}&693.24&0.38&410.01&0.38 \\
            \hline
            \textbf{RedBelly with resets\cite{BaumeisterEJSV24}} & & & & & & & & & & & & & & & & & & & \\
                rb-2x\_reset&49&28&\cmark&&&&&&&&&TO&-&TO&-&45.52&0.17&\textbf{37.25}&\textbf{0.15} \\
                rb-floodMin\_V0&10&7&\cmark&&&&&&&&&0.13&0.05&\textbf{0.08}&\textbf{0.03}&0.15&0.05&0.08&0.03 \\
                rb-floodMin\_V1&10&7&\cmark&&&&&&&&&0.12&0.05&\textbf{0.08}&\textbf{0.03}&0.15&0.05&0.08&0.03 \\
                rb-RelBrd\_V1&7&4&\cmark&&&&&&&&&0.11&0.05&\textbf{0.07}&\textbf{0.03}&0.11&0.05&\textbf{0.07}&\textbf{0.03} \\
                rb-reset\_V0&47&26&\cmark&&&&&&&&&TO&-&TO&-&\textbf{43.08}&\textbf{0.17}&80.60&0.19 \\
                rb-reset\_V1&47&26&\cmark&&&&&&&&&TO&-&TO&-&\textbf{145.23}&\textbf{0.21}&236.44&0.22 \\
                rb-simple-2x\_reset\_V0&43&21&\cmark&&&&&&&&&16.31&0.38&11.52&0.32&1.32&0.05&\textbf{1.28}&\textbf{0.04} \\
                rb-simple-2x\_reset\_V1&43&21&\cmark&&&&&&&&&736.52&24.97&1138.11&31.25&3.77&0.08&\textbf{5.47}&\textbf{0.07} \\
                rb-simple-reset\_V0&39&19&\cmark&&&&&&&&&275.26&11.86&334.97&13.89&\textbf{2.15}&\textbf{0.05}&2.27&0.04 \\
                rb-simple-reset\_V1&39&19&\cmark&&&&&&&&&574.49&18.94&239.67&13.16&\textbf{2.12}&\textbf{0.05}&3.66&0.04 \\
            \hline
        \end{tabular}
    }
\end{landscape}